\documentclass[10pt,a4paper,useAMS,usenatbib]{article}
\usepackage{jcappub}
\usepackage[english]{babel}

\usepackage{fancyhdr}
\usepackage{amsfonts}
\usepackage{amsmath}
\usepackage{amssymb}
\usepackage{multicol}
\usepackage{layout}
\usepackage{graphicx}
\usepackage{epstopdf}

\usepackage{times}
\usepackage{natbib}

\newif\ifAMStwofonts
\AMStwofontstrue

\title{The clustering of baryonic matter. II: halo model and hydrodynamic simulations}

\author[a,b]{C. Fedeli,}
\author[c]{E. Semboloni,}
\author[c]{M. Velliscig,}
\author[c]{M. Van Daalen,}
\author[c]{\\J. Schaye,}
\author[c]{and H. Hoekstra}

\affiliation[a]{INAF - Osservatorio Astronomico di Bologna, via Ranzani 1, 40127 Bologna, Italy}
\affiliation[b]{Department of Astronomy, University of Florida, 211 Bryant Space Science Center, Gainesville, FL 32611}
\affiliation[c]{Leiden Observatory, Leiden University, PO Box 9513, 2300 RA Leiden, the Netherlands}

\emailAdd{cosimo.fedeli@oabo.inaf.it}

\abstract{We recently developed a generalization of the halo model in order to describe the spatial clustering properties of each mass component in the Universe, including hot gas and stars. In this work we discuss the complementarity of the model with respect to a set of cosmological simulations including hydrodynamics of different kinds. We find that the mass fractions and density profiles measured in the simulations do not always succeed in reproducing the simulated matter power spectra, the reason being that the latter encode information from a much larger range in masses than that accessible to individually resolved structures. In other words, this halo model allows one to extract information on the growth of structures from the spatial clustering of matter, that is complementary with the information coming from the study of individual objects. We also find a number of directions for improvement of the present implementation of the model, depending on the specific application one has in mind. The most relevant one is the necessity for a scale dependence of the bias of the diffuse gas component, which will be interesting to test with future detections of the Warm-Hot Intergalactic Medium. This investigation confirms the possibility to gain information on the physics of galaxy and cluster formation by studying the clustering of mass, and our next work will consist of applying the halo model to use future high-precision cosmic shear surveys to this end.}

\begin{document}
\maketitle

\section{Introduction}\label{sct:introduction}

A number of cosmological experiments are being developed and planned that will allow to measure the growth of cosmic structure with exquisite precision (e.g., LSST \cite{LS09.1} and \emph{Euclid}\footnote{\texttt{www.euclid-ec.org/}} \cite{LA11.1}). Different surveys are going to employ different observational probes (e.g., cosmic shear, Baryon Acoustic Oscillation, redshift-space distortions, and so forth), however they will all have in common the need for an excellent characterization of statistical uncertainties. In this situation, it becomes increasingly important to have good control over the systematics, both observational and theoretical, affecting different cosmological observables. Theoretical systematics stem from our inability to understand and/or model the quantity that is being observed. For instance, in the case of galaxy clustering, one important systematic is given by galaxy evolution, which determines the bias with respect to the underlying Dark Matter (DM henceforth) distribution. In the case of cosmic shear the most relevant systematic is given by the nonlinear clustering of DM and baryons.

Cosmic shear is obtained by correlating the average distortion of galaxy images induced by gravitational lensing at different positions on the sky, and it effectively measures the projection of the \emph{total} matter power spectrum. Because cosmic shear is a projected quantity, it combines different physical scales at different redshifts, and hence for its interpretation it is important to be able to understand and correctly model even rather small, nonlinear scales. Future wide-field imaging surveys, especially the ESA mission \emph{Euclid}, will have enough statistical power to measure the full matter clustering with the astounding precision of a few percent \cite{LA11.1}.

The nonlinear part of the DM power spectrum is relatively well understood. Apart from perturbative approaches, which can be pushed up to spatial frequencies of a few tenths $h$ Mpc$^{-1}$, the full DM clustering can be captured down to small scales by numerical $N-$body simulations. DM clustering is driven only by gravity, which is a relatively simple physical process. Thus, despite widely different numerical implementations, there is a relatively fair consensus on what the nonlinear DM power spectrum looks like \cite{PE94.1,SM03.1,TA12.1} (see however the discussion in ref. \cite{SM14.1}). The same is not true when baryons are included in the picture. Physical processes such as preheating, radiative gas cooling, star formation, supernova feedback, metal enrichment, and AGN activity commonly originate on scales that are too small to be resolved by cosmological hydrodynamic simulations. As a consequence, sub-grid parametrizations and approximations need to be used, often based on incomplete knowledge of the physical process at hand. The literature shows that different sub-grid prescriptions for different physics lead to widely different clustering properties for both DM and baryons. For instance, the effect of cooling and star formation is to increase the total matter power spectrum by $\sim 5-50\%$ for $k\sim 5 h$ Mpc$^{-1}$, while the inclusion of AGN feedback can decrease the clustering power by $\sim 30\%$ on the same scales \cite{JI06.1,RU08.2,CA11.1,VA11.1,FE11.1}.

Unfortunately, the run time of cosmological hydrodynamic simulations is such that a systematic exploration of the astrophysical parameter space is not feasible, especially if one also wants to investigate the degeneracies with cosmology. Thus, it is worth considering different modeling approaches. In a companion paper (\cite{FE14.1}, Paper I henceforth) Fedeli presented a generalization of the well known halo model in order to describe the clustering of hot gas and stars, in addition to the DM clustering. The model involves three different density profiles for the three mass contributions, and includes a diffuse gas component in order to take into account baryons that have not accreted into bound structures. While having in common with hydrodynamic simulations the dependence on a number of free parameters, this model allows for much faster calculations, and it thus constitutes an excellent complement to expensive numerical simulations. Previous efforts in the same direction, usually at a more simplified level, can be found in refs. \cite{SE11.1,SE13.1,ZE13.1}.

Obviously, like many Semi-Analytic Methods (SAMs henceforth) do, our model also relies on several assumptions and approximations. Here we make use of two simulations from the OverWhelmingly Large Simulation (OWLS) project \cite{SC10.1} in order to investigate the validity of these approximations, and discuss if and how the model can be improved in the future. At the same time, we emphasize what can be learned about the nonlinear mass clustering that cannot be learned with hydrodynamic simulations, and also study the inverse problem of what kind of information the model can return on galaxy and cluster formation physics, provided that we have access to a high-precision measurement of the total matter power spectrum, or any of its components.

The rest of the paper is organized as follows. In Section \ref{sct:simulations} we briefly describe the two simulations that have been used here. In Section \ref{sct:fractions} we discuss the mass fractions measured in the simulations and their modeling. In Section \ref{sct:spectra} we compare the power spectra of each mass component produced by the simulations to the results of the SAM, and draw some information on the abundance and distribution of baryons within individual structures. Our conclusions are discussed in Section \ref{sct:conclusions}. For consistency, we used the cosmological parameters of the OWLS project, although several of them are slightly offset with respect to the latest \emph{Planck} results \cite{PL13.1}. They correspond to a flat $\Lambda$CDM cosmology with $\Omega_{\mathrm{m},0} = 0.238$, $\Omega_{\mathrm{b},0} = 0.042$, $h\equiv H_0/(100h\mathrm{\;km\;s}^{-1}\mathrm{\;Mpc}^{-1})=  0.730$, $\sigma_8=0.740$, and $n=0.951$, where $n$ is the scalar spectral index.

\section{Simulations}\label{sct:simulations}

The simulations of the OWLS project were run with a modified version of the Smoothed Particle Hydrodynamics (SPH) code \texttt{Gadget}3 (last described in ref. \cite{SP05.2}). We summarize only the main features here, redirecting the interested reader to the paper by Schaye and collaborators \cite{SC10.1} for further details. The simulations we used were run in a box with a comoving side of $100h^{-1}$ Mpc and contain $512^3$ DM particles and $512^3$ baryon particles, the latter being either gas or stars. Particle masses are thus $4.06\times 10^8h^{-1}M_\odot$ for DM and $8.66\times10^{7}h^{-1}M_\odot$ for baryons. The latter however is not a constant throughout the simulation because of the mass transfer between gas and stars. Gravitational softening was set to be $1/25$ times the mean comoving interparticle separation, with a maximum physical scale of $2h^{-1}$ kpc. All simulations were started at $z=127$, with CAMB-generated initial conditions being evolved there through the Zel'Dovich approximation.

The OWLS project has recently been expanded with new simulations having larger sizes and the same physics with respect to those described above \cite{LE13.1}. These simulations were not available during the development of this work, and thus they have not been included in this discussion. Moreover, the new simulations were evolved in a WMAP$-7$ cosmology, while the old ones that we used were evolved in a WMAP$-3$ cosmology. Although this change seems to have a mild effect on the total mass profile of structures \cite{VE14.1}, it is possible for the effect to be more significant on the clustering of gas and stars when taken individually.

In addition to two hydrodynamic simulations, we also considered the baryon-free simulation, that has been labeled as DMONLY. The two hydrodynamic runs are labeled REF and AGN, respectively. The REF simulation represents a standard hydro simulation including radiative cooling, star formation, supernova winds and stellar evolution, but no AGN feedback. The latter has been included, as per the name, in the AGN simulation. The AGN activity can heat up and displace large amounts of gas, and the AGN run has been argued to be the most realistic amongst the OWLS \cite{MC10.1,MC11.1,LE13.1,VE14.1}. Gravitationally bound structures were identified using a friends-of-friends algorithm, with a linking length equal to $0.2$ times the mean inter-particle separation. In order to have well-established mass density profiles, we required any object to be composed by at least $1,000$ particles in order for it to be included in the analysis that follows. It is noteworthy that all simulations have been run with the same initial conditions, meaning that individual structures in one simulation can be matched to their counterparts in another simulation. As discussed in Paper I, this allows one to link the masses of different structure components in the hydrodynamic simulations to that of the corresponding structure in the baryon-free case. We refer to this mass as the \emph{equivalent mass} henceforth. Note that the equivalent mass is not observable, however its difference with respect to the total mass (which is observable) is slight.

\section{Mass fractions}\label{sct:fractions}

\begin{figure}
	\centering
	\includegraphics[width=0.49\hsize]{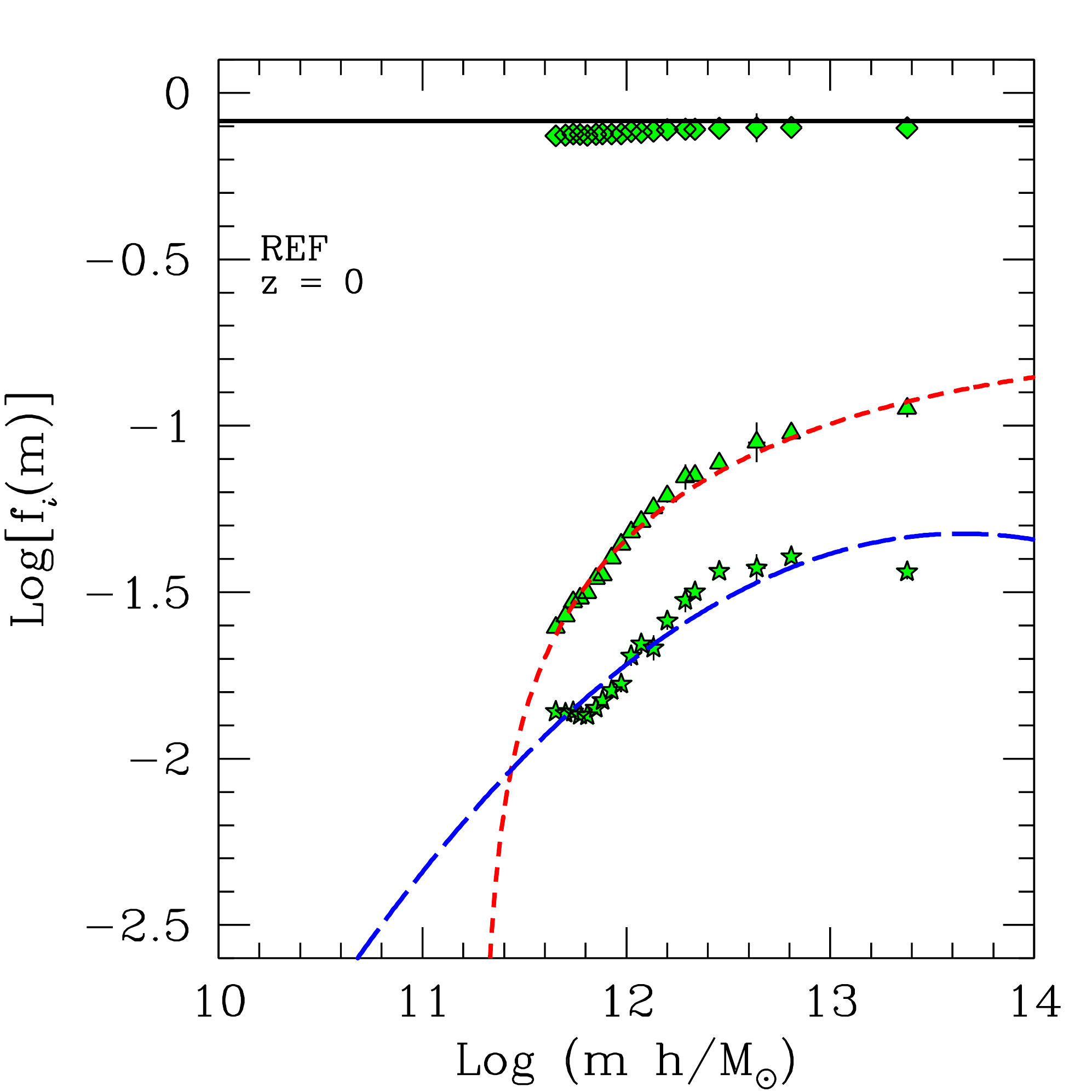}
	\includegraphics[width=0.49\hsize]{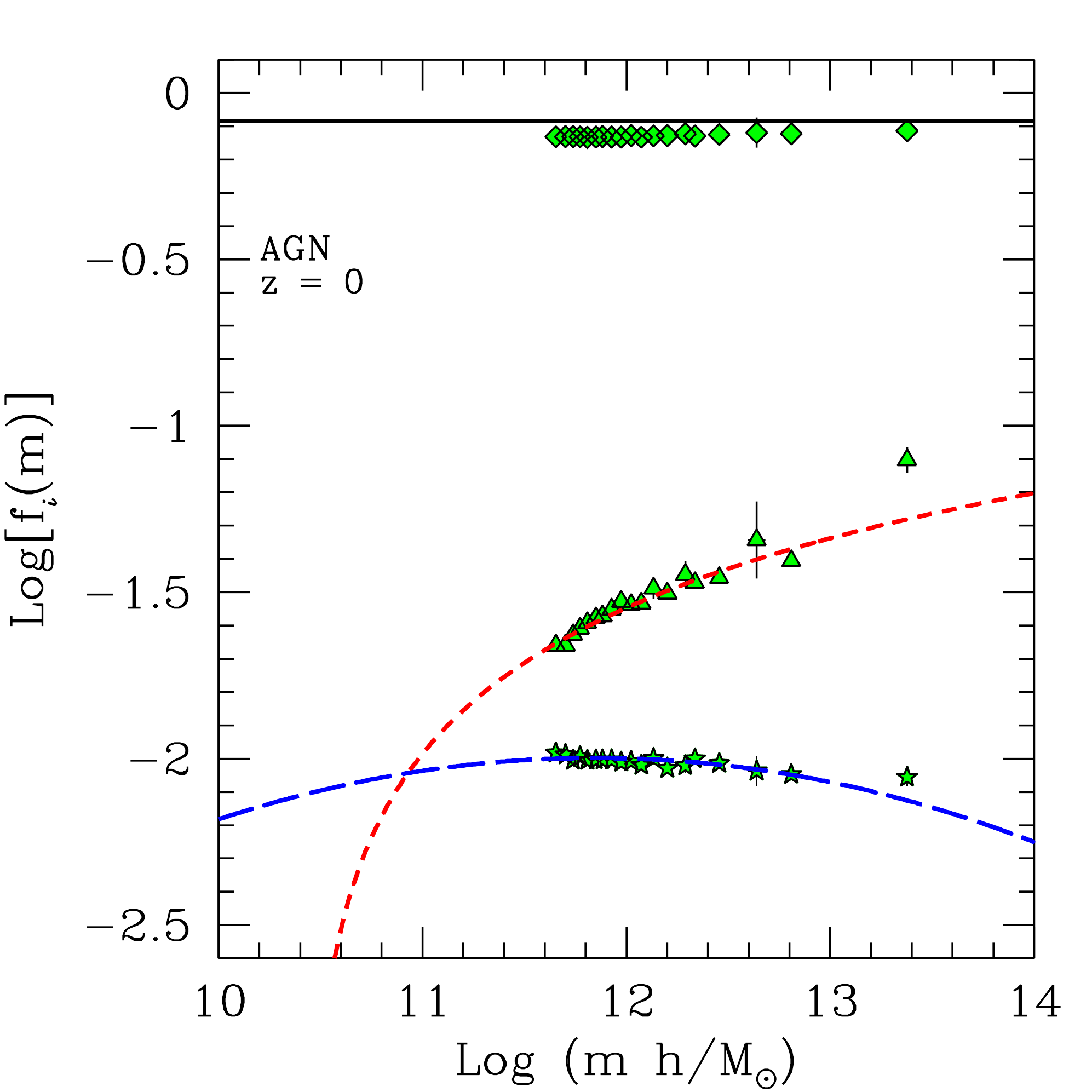}
	\caption{The mass fractions of different matter components measured in the REF (left panel) and AGN (right panel) runs as a function of the structure equivalent mass: DM (diamonds), gas (triangles), and stars (stars). Vertical bars display the profile-to-profile scatter around the mean stacked profiles, however they are visible only for a few points. The red and blue dashed curves show the best fit to the gas and stellar mass fractions, respectively, while the black solid line displays the quantity $1-\Omega_{\mathrm{b},0}/\Omega_{\mathrm{m},0}$ (see the text for more details).}
	\label{fig:fractions}
\end{figure}

The first piece of information that we need to know in order to use the SAM developed in Paper I, is the contribution of each matter component to the total mass of a certain structure. Thus, we measured mass fractions in the two hydrodynamic OWLS simulations considered in this work by stacking all structures within a certain equivalent mass bin and then integrating the resulting stacked profile out to the corresponding virial radius\footnote{As discussed in Paper I, we define the virial radius as the radius of the sphere encompassing a mean overdensity of $200$ times the comoving average density of the Universe.}. We chose the equivalent mass bins such that each bin hosts approximately the same number of structures, so that the error bars reported in this work represent the actual scatter between objects in a bin, rather than Poisson fluctuations. Baryon fractions measured in the REF and AGN simulations are shown as green triangles (gas) and stars (stellar) in Figure \ref{fig:fractions}. Being integrated quantities, these fractions are very stable, and the scatter is small enough to be visible for only a handful of points. The red and blue dashed lines show the best fits to the mass fractions of gas and stars, adopting the functional forms introduced in Paper I (an error function and a Gaussian, respectively). In the interest of reducing the number of free parameters, the stellar fraction is always normalized to reproduce the mean universal stellar density measured in the simulations. For the same reason, the SAM also assumes that the bulk properties of DM halos are not affected by baryonic physics, so that $f_\mathrm{DM} = 1-\Omega_{\mathrm{b},0}/\Omega_{\mathrm{m},0}$ at all masses. This value is displayed in Figure \ref{fig:fractions} with a black solid line, while the DM fractions actually measured in simulations are represented by green diamonds.

As one might expect, there are a number of significant differences between the two hydrodynamic simulations considered here, more so at the high-mass end. Both the stellar and the gas mass fractions are significantly suppressed in the AGN run compared to the REF run, which is likely due to the substantial AGN feedback inhibiting star formation and ejecting large amounts of gas out of bound structures. Moreover, both the stellar and gas mass fractions display a much less pronounced variation with mass in the AGN simulation than in the REF one. This suggests that AGN feedback is more effective at suppressing star formation efficiency in denser environments. We also note that in both simulations the DM mass fraction does not show any significant trend with equivalent mass, although in the REF simulation it slightly increases with mass. Overall, the DM mass fraction is somewhat lower than the universal value we assumed, slightly more so in the AGN run. The implication of this is that the accretion of DM onto gravitationally bound structures is not completely independent of the baryons, rather it gets slightly delayed in the presence of a feedback (with a stronger feedback causing more delay). It is curious that this modification is almost independent of the mass of the structure, however Velliscig and collaborators \cite{VE14.1} recently made use of the extended mass range allowed by the new OWLS simulations to confirm that the DM fraction does indeed grow somewhat with increasing mass. Because the SAM works under the assumption that all DM is locked into bound structures, we cannot simply lower the DM fraction by a constant at all masses $\lesssim 10^{13.5}h^{-1}M_\odot$, since this would alter the universal DM density. Moreover, the introduction of a different functional form for $f_\mathrm{DM}(m)$ would be artificial in the low mass regime, since it is not supported by the simulations, nor by physical arguments. This, together with the need to keep the number of free parameters limited and the fact that the model results do not change appreciably, motivated us to stick to the original choice.

It is worth noting that the mass range covered by the simulations used here is rather limited. On small scales it is limited by the resolution, as we require that structures are identified by at least $1,000$ particles, meaning $m \gtrsim 4\times 10^{11}h^{-1}M_\odot$. On large scales the limit is given by the simulation size, that does not yield sufficient statistics for masses larger than $m\sim 10^{14}h^{-1}M_\odot$. This limitation implies that we have to rely heavily on extrapolation in order to infer the mass fractions and the density profiles at low and high masses. As is evident from Figure \ref{fig:fractions}, this might be a problem especially for the stellar fraction in the AGN run, which is virtually independent of mass for the mass range accessible to these simulations. Interestingly, Budzynski and collaborators \cite{BU13.1} recently also found a rather flat observed stellar mass fraction trend with total mass, although their mass range covered the groups and cluster scales, and is thus higher than ours. In any case, the flatness of the stellar fraction implies high uncertainties in its extrapolation at high and low masses, which is bound to have an impact on the resulting stellar power spectrum. This will be explored in more detail in Section \ref{sct:stars}.

\section{Matter power spectra}\label{sct:spectra}

\subsection{Dark matter}

\subsubsection{Baryon-free case}

We first discuss the DM power spectrum in the case of no baryons, thus setting $f_\mathrm{g} = f_\star = 0$ in the SAM and considering the reference DMONLY simulation. In the left panel of Figure \ref{fig:darkMatterPower_COMPARISON} we show the ratio of the DM power spectrum measured in the DMONLY run to its counterpart estimated through the halo model. Blue circles show the original implementation of the SAM, illustrated in Paper I. As can be seen, the halo model tends to underestimate the DM power by $\sim 15\%$ at the transition between the $1-$halo and the $2-$halo contributions (refer to the right panel in Figure \ref{fig:darkMatterPower_COMPARISON}) and by up to $\sim 35\%$ at small scales ($k\gtrsim 3h$ Mpc$^{-1}$). 

The lack of power at intermediate scales is a well known shortcoming of the halo model \cite{RE02.2,HU03.2,FE10.1}, and it stems from two facts: firstly, as quasi-linear scales are approached one should replace the linear DM power spectrum in the $2-$halo term by some higher-order approximation, possibly arising from perturbation theory \cite{CR06.1,CR06.2,VA13.1}, in order to take into account the coupling of Fourier modes that is not appropriately captured by the onset of the $1-$halo term \cite{BE02.2}. Secondly, the simple linear biasing approximation that is implicitly adopted in the halo model illustrated in Paper I also ceases to be valid at quasi-linear scales. As a consequence, one should include in the formalism higher-order bias factors, effectively obtaining a scale dependent bias of DM halos \cite{VA13.2}. The fact that halo bias has some sort of scale dependence has been known for some time \cite{CO05.1,CR09.2}, and many authors have proposed different prescriptions for its modeling \cite{SE01.1,CO05.1,SE05.2,SC06.1,GU07.1,HU07.1,MU12.1}. In \cite{SM07.1} Smith, Scoccimarro, \& Sheth found on theoretical grounds that the scale dependence of the halo bias has to be a function of mass, which has been observationally verified in \cite{CR09.2}; these authors found that the scale dependence of galaxy bias is a function of both galaxy type and luminosity. The effect of the non-linear coupling of Fourier modes is customarily mitigated by heuristically modifying the $2-$halo contribution in order to include a halo exclusion term, taking into account the fact that DM halos cannot overlap with each other \cite{SH01.2,YA03.1,SM11.2}. However, this leaves the question of the scale dependence of the halo bias open, which still needs to be calibrated against numerical simulations in order to properly describe the clustering of matter on quasi-linear scales \cite{TI05.1,VA13.2}. Since the halo exclusion does not eliminate the need for a numerical calibration, and at the same time it substantially complicates the calculation of the $2-$halo term, we decided instead to encapsulate all the power needed on  quasi-linear scales in the scale dependence of the halo bias. After some experimentation, we found that the following substitution produces an acceptable result:

\begin{figure}
	\centering
	\includegraphics[width=0.49\hsize]{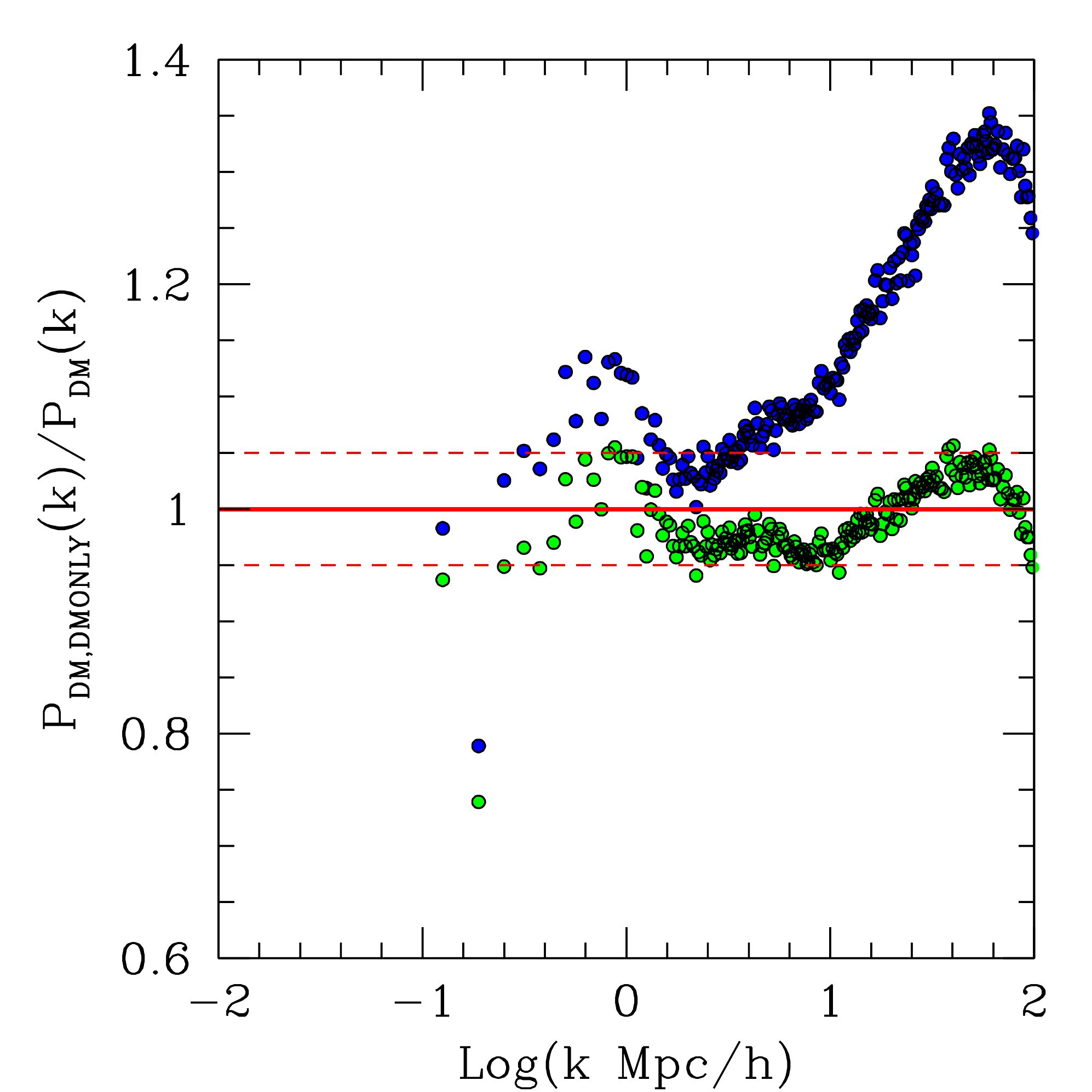}
	\includegraphics[width=0.49\hsize]{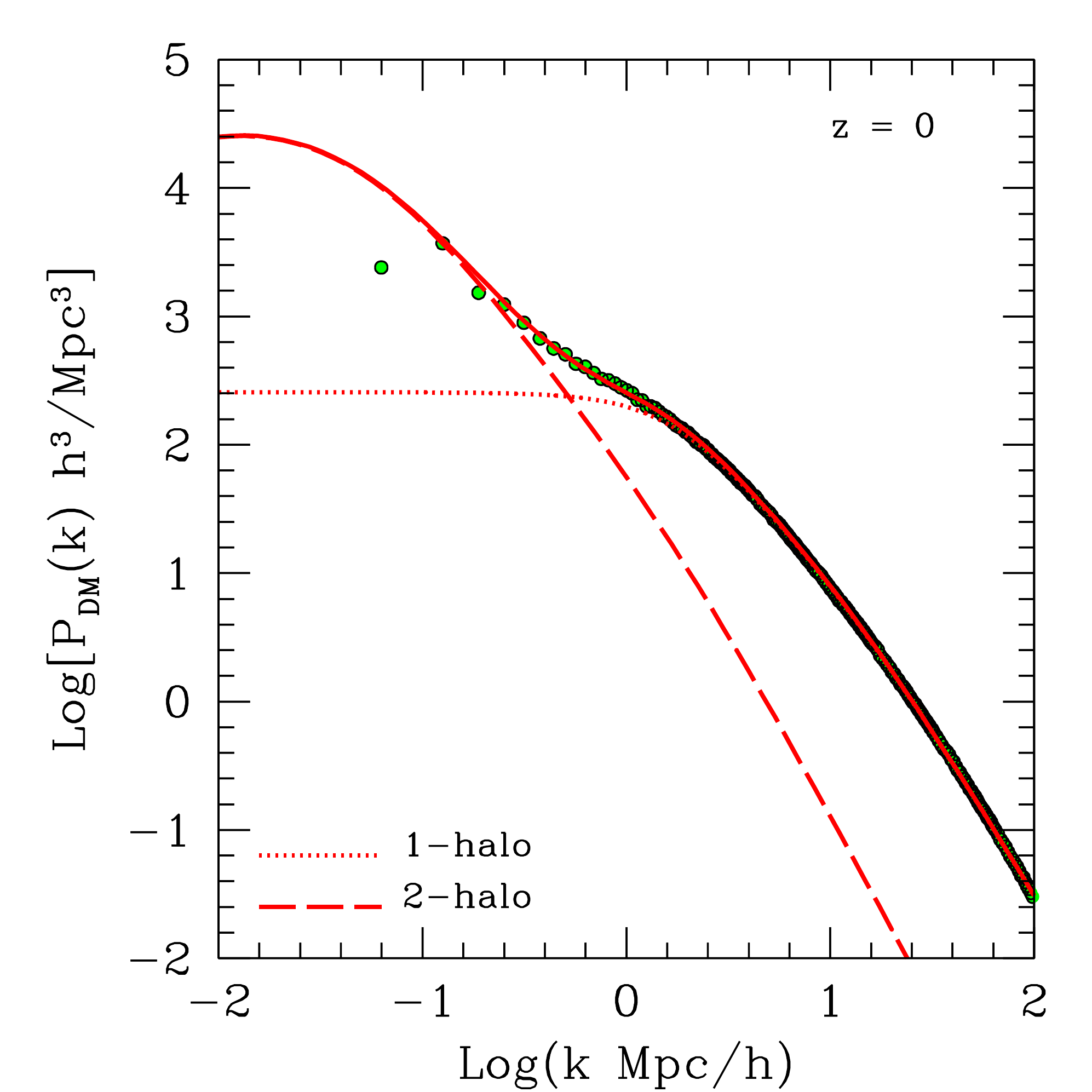}
	\caption{\emph{Left panel}. The ratio of the DM power spectrum measured in the DMONLY run to the same quantity estimated through the SAM developed in Paper I. Blue circles refer to the original implementation of the model, while green circles refer to the improved implementation featuring a scale dependent halo bias and an adjusted concentration-mass relation for DM halos (see the text for details). The horizontal dashed lines illustrate a deviation of $\pm 5\%$ from the semi-analytic power spectrum. \emph{Right panel}. Direct comparison of the DM power spectrum measured in the DMONLY simulation and its halo model counterpart with improved implementation (red solid line). The red dotted and dashed lines illustrate the $1-$halo and $2-$halo contributions, as labeled.}
	\label{fig:darkMatterPower_COMPARISON}
\end{figure}

\begin{equation}\label{eqn:biasScale}
b(m)\rightarrow b(k,m) = b(m)\left( 1+\frac{k}{1h~\mathrm{Mpc}^{-1}} \right)^{0.27}\;.
\end{equation}
As illustrated by the green circles in Figure \ref{fig:darkMatterPower_COMPARISON}, the discrepancy with respect to the DMONY run is brought down from $\sim 15\%$ to $\sim 5\%$. Although this discrepancy is still somewhat large compared to the expected statistical errors from \emph{Euclid}, improving upon this is difficult because the scatter in the power measured in the DMONLY run within a fixed mode is also $\sim 5\%$. One possible approach to follow in future observational studies is to consider the scale dependence of the halo bias as a nuisance, and marginalize over it when inferring cosmology and/or baryonic physics. Note that Eq. (\ref{eqn:biasScale}) does not look like other prescriptions for a scale-dependent bias, nor is it supposed to.

The lack of power on small scales is much easier to address, as it results from the substantial oversimplification of the average DM density profile that we included in the SAM. Real DM halos tend to be elongated, substructured, and to have some scatter in the concentration for a given mass, none of which has been included in the present implementation of the halo model. As has been shown in \cite{GI10.1}, including the substructure contribution to the matter power and convolving the average DM density profile with a concentration distribution (thus effectively realizing a \emph{stochastic} density profile rather than a \emph{deterministic} one), can increase the power spectrum by up to $\sim 30\%$ on small scales, which is about the discrepancy we found. In the interest of simplicity, this has been modeled by adopting a new concentration-mass relation that reads

\begin{equation}
c_\mathrm{DMONLY}(m) = 13~\left(\frac{m}{10^{12}h^{-1}M_\odot}\right)^{-0.105}~.
\end{equation}
The normalization is $\sim 18\%$ higher than the fiducial one illustrated in Paper I, while the slope has been made slightly steeper (it was $0.100$ before). As can be seen in Figure \ref{fig:darkMatterPower_COMPARISON}, this reduces the deviation with respect to the DMONLY run also to $\sim 5\%$. Obviously, this modified concentration-mass relation is now not expected to be a good fit to the individual profiles measured in the numerical simulations.

\subsubsection{Impact of baryons}

\begin{figure}
	\centering
	\includegraphics[width=0.49\hsize]{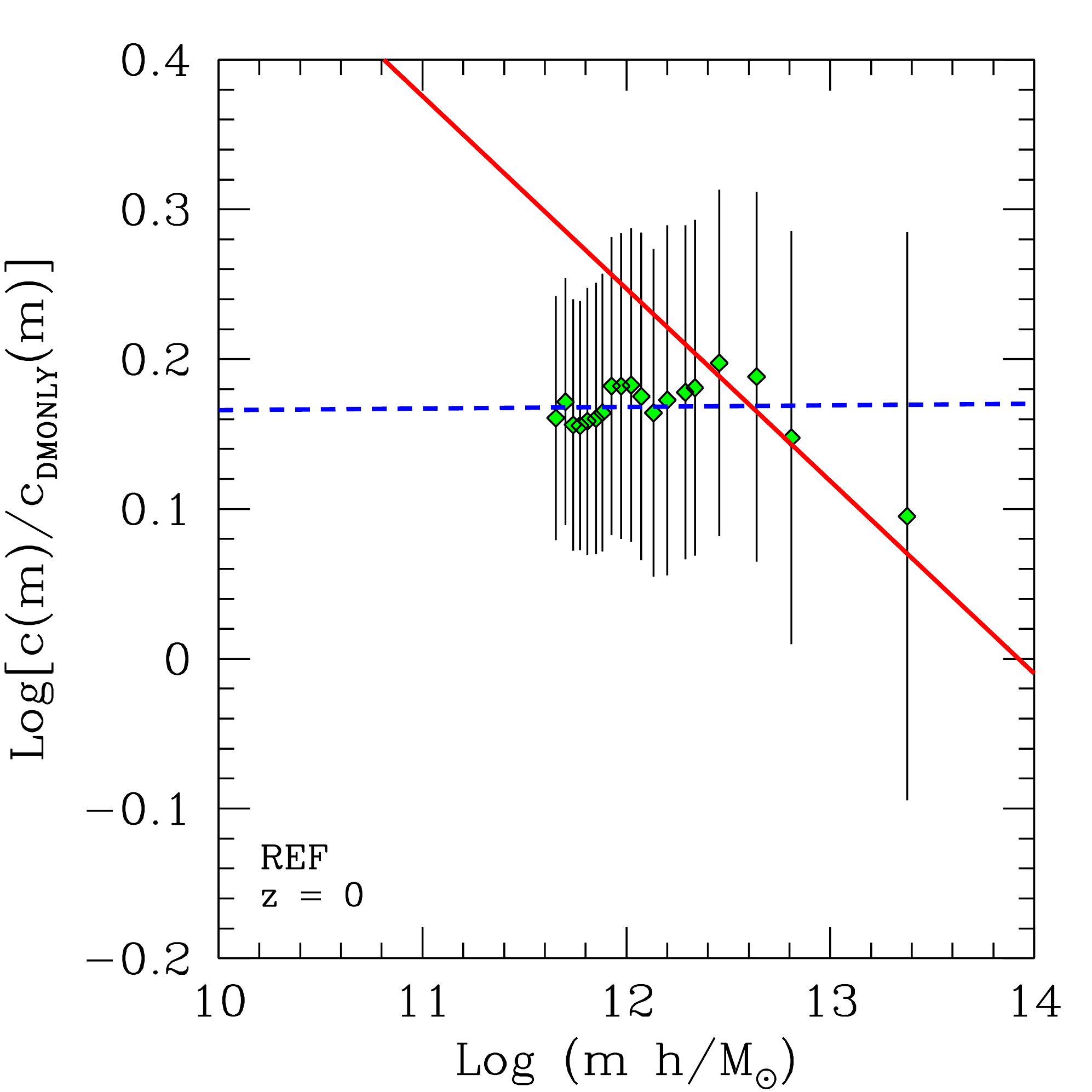}
	\includegraphics[width=0.49\hsize]{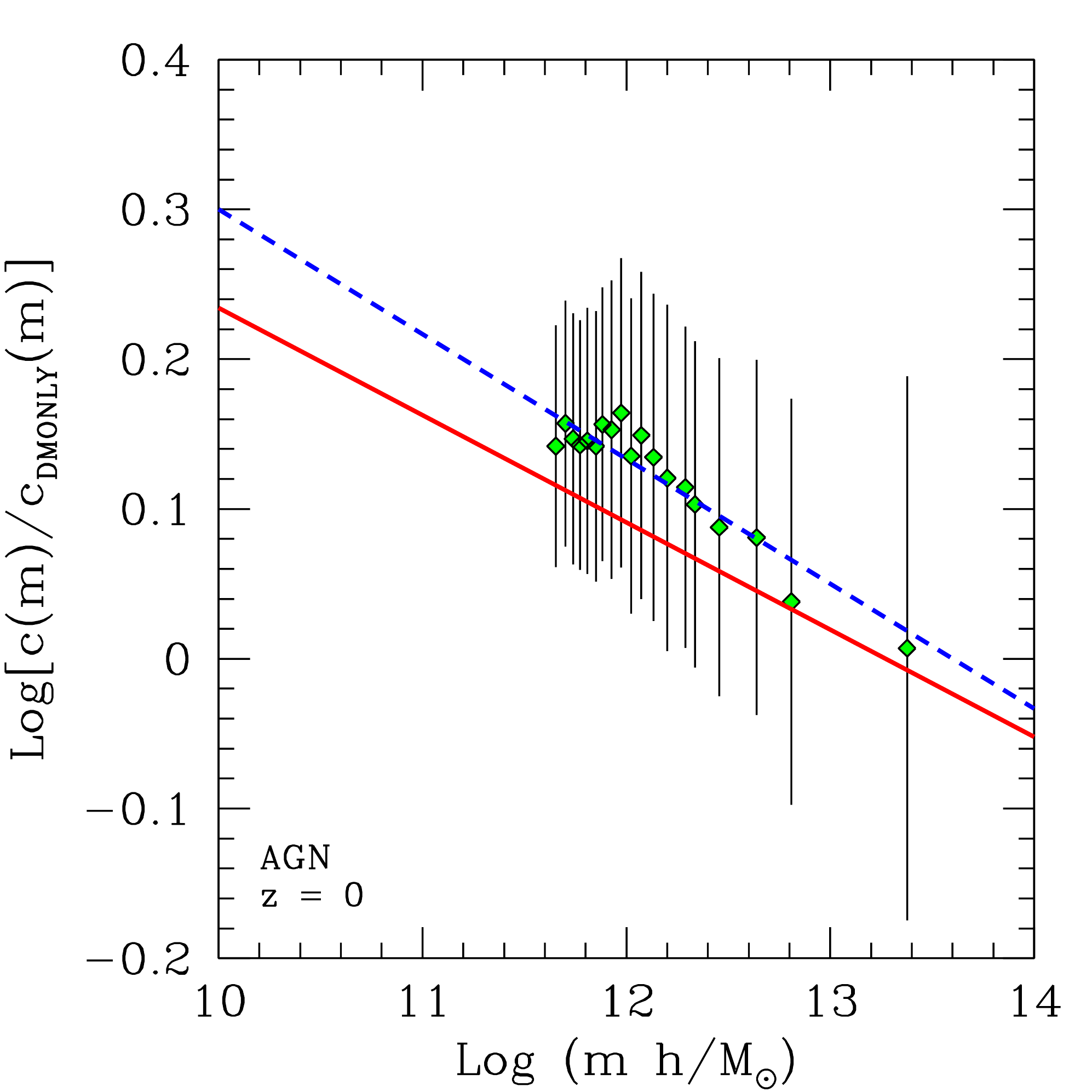}
	\caption{The ratio of the DM halo concentration for a given equivalent mass to its counterpart in the DMONLY simulation. Diamonds represent results for the REF run (left panel) and AGN run (right panel). Halos have been stacked within mass bins so that about the same number is present within each bin. Vertical bars show the scatter around the mean stacked profiles. The blue dashed lines show the best fit power-laws, while the red solid lines show the power-laws giving the best fitting power spectra (see text for details).}
	\label{fig:darkMatter_c}
\end{figure}

We now examine the behavior of DM clustering when baryons modeled with different physics are present. We first look into the behavior of individual DM halos and use the SAM to see how this propagates into the DM power spectrum, and then take the opposite approach. Specifically, we fit the stacked DM profiles in different mass bins (the same bins as adopted in Figure \ref{fig:fractions}) with a NFW function \cite{NA96.1} and retrieve the halo concentrations. In Figure \ref{fig:darkMatter_c} we show the ratios of the DM halo concentrations in the REF and AGN runs to their counterparts in the reference DMONLY simulation. Vertical bars represent the $1-\sigma$ scatter around the ratio, propagated from the scatter around each mean concentration. 

A few facts can be extracted from Figure \ref{fig:darkMatter_c}. Firstly, we note that there is a rather large scatter (a factor of $\sim 2-2.5$) in the concentration ratio at a fixed halo mass. Secondly, the concentrations in the REF run are systematically higher than those in the DMONLY run by $\sim 40-50\%$. This can be interpreted as the contraction effect that the cooling of gas and the formation of stars exert on the DM distribution. Concentrations are also increased by about the same factor in the AGN run at low masses, however the concentration ratio tends to about unity for high masses. Anyway, the large scatter in the concentration ratio implies that the trend of this ratio with mass cannot be easily ascertained. A similar study on how the DM structure of halos evolves in the presence of baryons has been performed by Duffy and collaborators \cite{DU10.1} on a subset of the OWLS simulations which includes our own. However, comparison with that work is hampered by two factors. First, they used a different structure definition (as reported in Paper I, we consider mean overdensities of $\Delta=200$ times the average comoving matter density, while the authors of \cite{DU10.1} use $\Delta = \Delta_\mathrm{v}(z)/\Omega_\mathrm{m}(z)$, where $\Delta_\mathrm{v}(z)$ is the virial overdensity according to the spherical collapse model). Second, they imposed a more restrictive cut on the number of particles composing a certain structure compared to our own ($10,000$ particles within the virial radius versus $1,000$ within $R_\Delta$). This greatly reduces the mass range available, so that a mean trend cannot be established in \cite{DU10.1}.

\begin{figure}
	\centering
	\includegraphics[width=0.49\hsize]{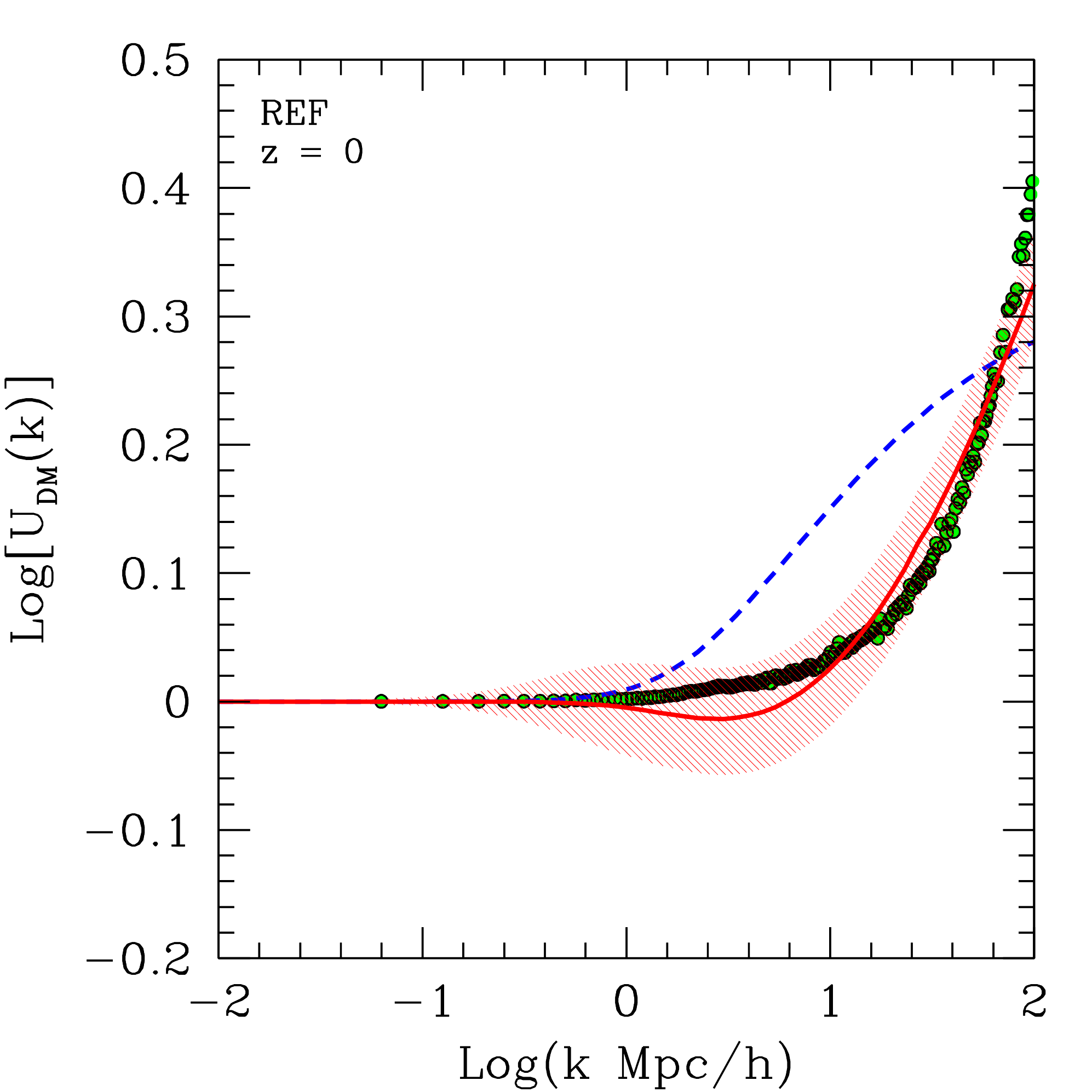}
	\includegraphics[width=0.49\hsize]{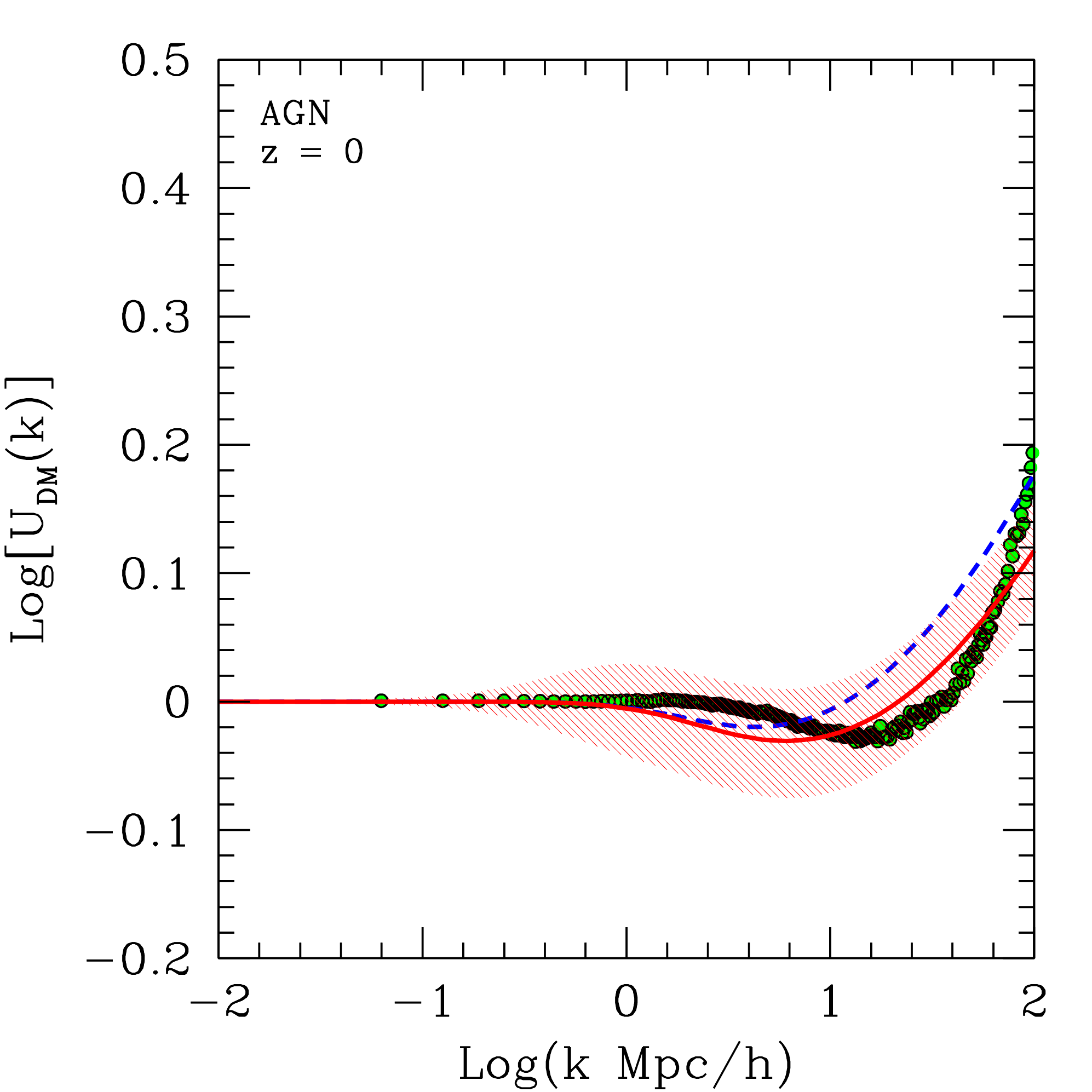}
	\caption{The ratio of the DM power spectra in the REF (left panel) and AGN (right panel) runs to the DM power spectrum in the DMONLY simulation (green circels). The blue dashed lines show the power spectra computed with the SAM adopting the best fit concentration-mass relations shown in Figure \ref{fig:darkMatter_c}, while the red solid lines show the best fitting power spectra produced by the SAM. The red shaded regions display the effect of a $\pm 10\%$ uncertainty on the DM halo mass function.}
	\label{fig:darkMatterPower_U_COMPARISON}
\end{figure}

The blue dashed lines shown in Figure \ref{fig:darkMatter_c} represent the best power-law fits to the concentration ratio as a function of mass. In Figure \ref{fig:darkMatterPower_U_COMPARISON} we show the DM power spectrum measured in the REF (left panel) and AGN (right panel) numerical runs, and compare it with the result of the SAM after feeding it with the best power-law fits for the concentration ratios. Here and in what follows we compare all spectra to the DM power spectrum in the DMONLY run, thus plotting $U_i(k)\equiv P_i(k)/P_\mathrm{DMONLY}(k)$. As already discussed in \cite{VA11.1}, the baryonic physics included in the REF run (mainly gas cooling and star formation) causes the DM power to increase steadily with increasing spatial frequency, reaching a factor of $\sim 2$ increment at $k \sim 50 h$ Mpc$^{-1}$. In the AGN run such an enhanced small-scale clustering can also be noticed (but only at the $\sim 10\%$ level). However, on larger scales ($k\sim 10-20 h$ Mpc$^{-1}$) the strong feedback from AGN causes a redistribution of the DM, so that the clustering is suppressed (also by $\sim 10\%$) with respect to the DMONLY run (see again \cite{VA11.1}). The halo model captures these features qualitatively, but because of the substantial scatter in the DM structure of individual halos, there are quantitative differences that we now proceed to discuss. 

As we have seen in Figure \ref{fig:darkMatter_c}, the DM halo concentrations in the REF simulations are increased by about the same factor at all masses with respect to the DMONLY run. This results in DM power being overestimated on intermediate scales, and underestimated at small scales, $k\gtrsim 50 h$ Mpc$^{-1}$. 
In the context of the halo model, this happens because the concentrations of high-mass structures (which dominate the clustering on intermediate scales) are being overestimated, while the concentrations of low-mass objects are being slightly underestimated. On the other hand, the SAM seem to work rather well for the AGN simulation for $k\lesssim 10 h$ Mpc$^{-1}$ (the DM power is underestimated by at most $\sim 5\%$ at $k \sim 3h$ Mpc$^{-1}$), while it overestimates DM clustering at smaller scales, although by a smaller amount than in the REF case. This can be also interpreted as an excess in the concentration of intermediate to low-mass structures.

In order to double-check this, we reversed the problem and fit the DM power spectra by releasing the amplitude and the slope of the concentration ratio-mass relations (see \cite{ZE13.1} for a similar approach, applied to the total matter power spectrum). The best fit spectra so obtained are shown as solid red curves in Figure \ref{fig:darkMatterPower_U_COMPARISON}, while the resulting power-laws are shown in Figure \ref{fig:darkMatter_c} with the same line type and color. As can be seen, the concentration ratios that best fit the DM power spectrum in the REF simulation change much more steeply with mass than before, thus implying a substantially higher concentration for structures with $m \lesssim 3\times 10^{12}h^{-1}M_\odot$. In the AGN case, the difference between the blue and red lines is less evident, but it still goes in the direction that we suspected: low-mass DM halos need to be less concentrated than a naive extrapolation of the concentration ratio-mass relation would suggest. It should be noted that the two power-laws depicted in each panel of Figure \ref{fig:darkMatter_c} cannot be easily distinguished by looking only at the DM structure of individual halos. This illustrates the power of having a physically based semi-analytic model to describe the clustering of matter: by having a high precision measurement of the (dark in this case) matter power spectrum, one can infer the average properties of individual structures, and on a range of masses much larger than that accessible to observations of individual structures only. This concept is reiterated also in what follows, in the more observationally significant cases where baryons are considered.

By looking closely at Figure \ref{fig:darkMatterPower_U_COMPARISON}, we can see that there are residual differences with respect to the numerical simulations, at the level of $\sim 5-10\%$. These are due to the fact that a single power law is probably not an accurate representation of the correlation between concentration ratios and mass. For instance, it is unlikely that DM halo concentrations continue to decrease indefinitely as mass is increased. More likely, they will reach a plateau somewhere between group and cluster mass scales, a regime that cannot be explored with the OWLS used in this work. This means that by considering the red lines in Figure \ref{fig:darkMatter_c} we are underestimating the compactness of high-mass structures, which results in the slight underestimation of the DM clustering power visible at $k\sim 3 h$ Mpc$^{-1}$. Mending this would require to add more free parameters to the model, which we preferred to avoid. This is justified by the fact that the uncertainty in the theoretical halo mass function is already more than enough to cover this discrepancy. The red shaded areas that we show in Figure \ref{fig:darkMatterPower_U_COMPARISON} represent the effect of a $\pm 10\%$ uncertainty on the mass function of DM halos, which is lower than the actual uncertainty \cite{TI08.1}. As can be seen, within this uncertainty $-$ provided it is not systematic $-$ the SAM is perfectly suited to reproduce the results of hydrodynamic numerical simulations.

Before concluding this Section, it is worth reminding the reader that the power-law we adopted to fit the trend with mass of the profile parameters (for both DM and baryons, which follow) has a constant pivot mass of $10^{12}h^{-1}M_\odot$ (see Paper I). This however may be not the optimal choice. Ideally, one should change the pivot mass for each parameter so as to minimize the covariance between the amplitude and the slope of the best-fit power-law. However, it is likely that this additional complication changes little the final result, so we leave it as a future improvement of the model. On a related note, we also stress that the results shown in Figures \ref{fig:darkMatter_c} and \ref{fig:darkMatterPower_U_COMPARISON} are only valid under the assumption that the NFW is a good fit to DM profiles. Similar cautionary considerations also apply to the other mass components.

\subsection{Gas}

\begin{figure}
	\centering
	\includegraphics[width=0.49\hsize]{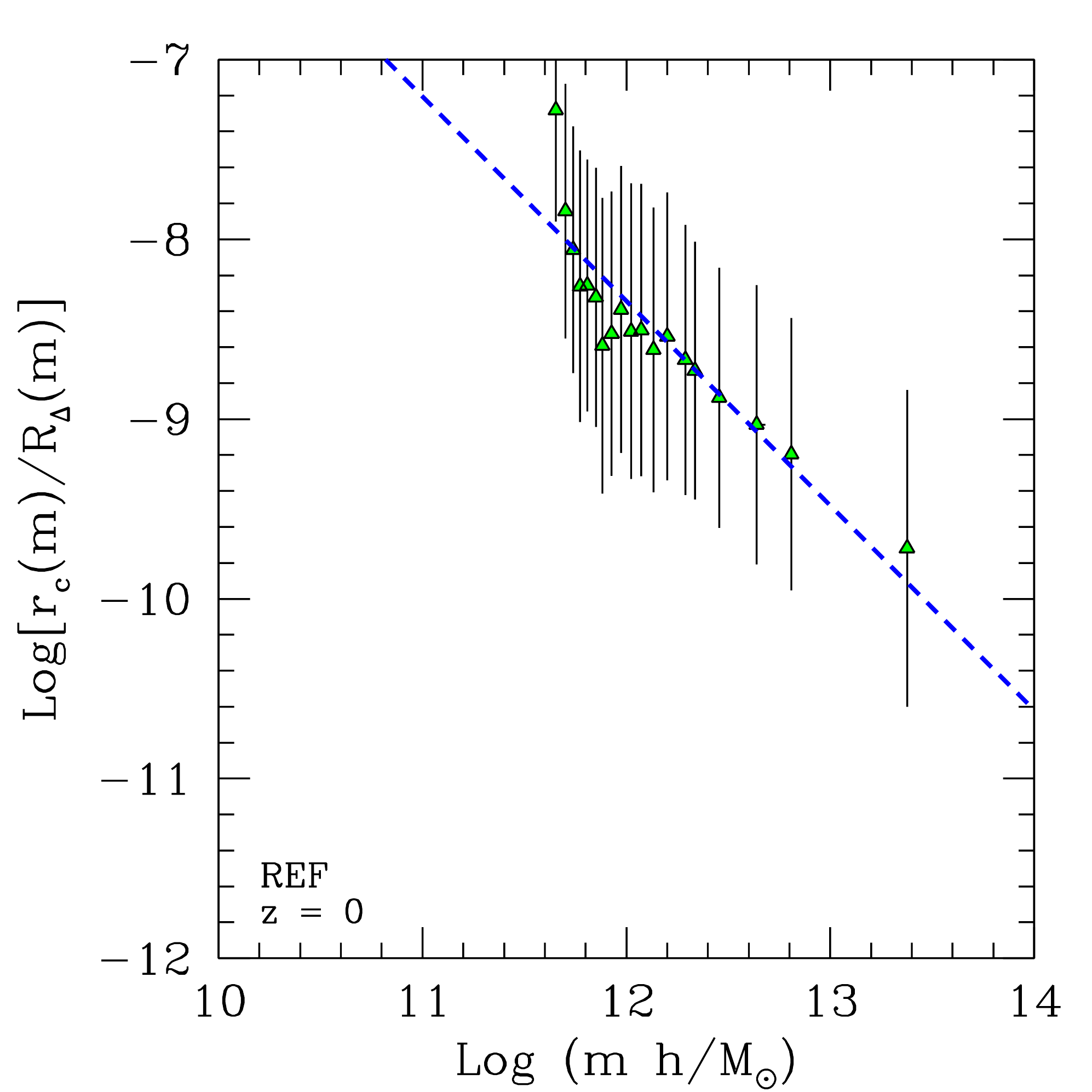}
	\includegraphics[width=0.49\hsize]{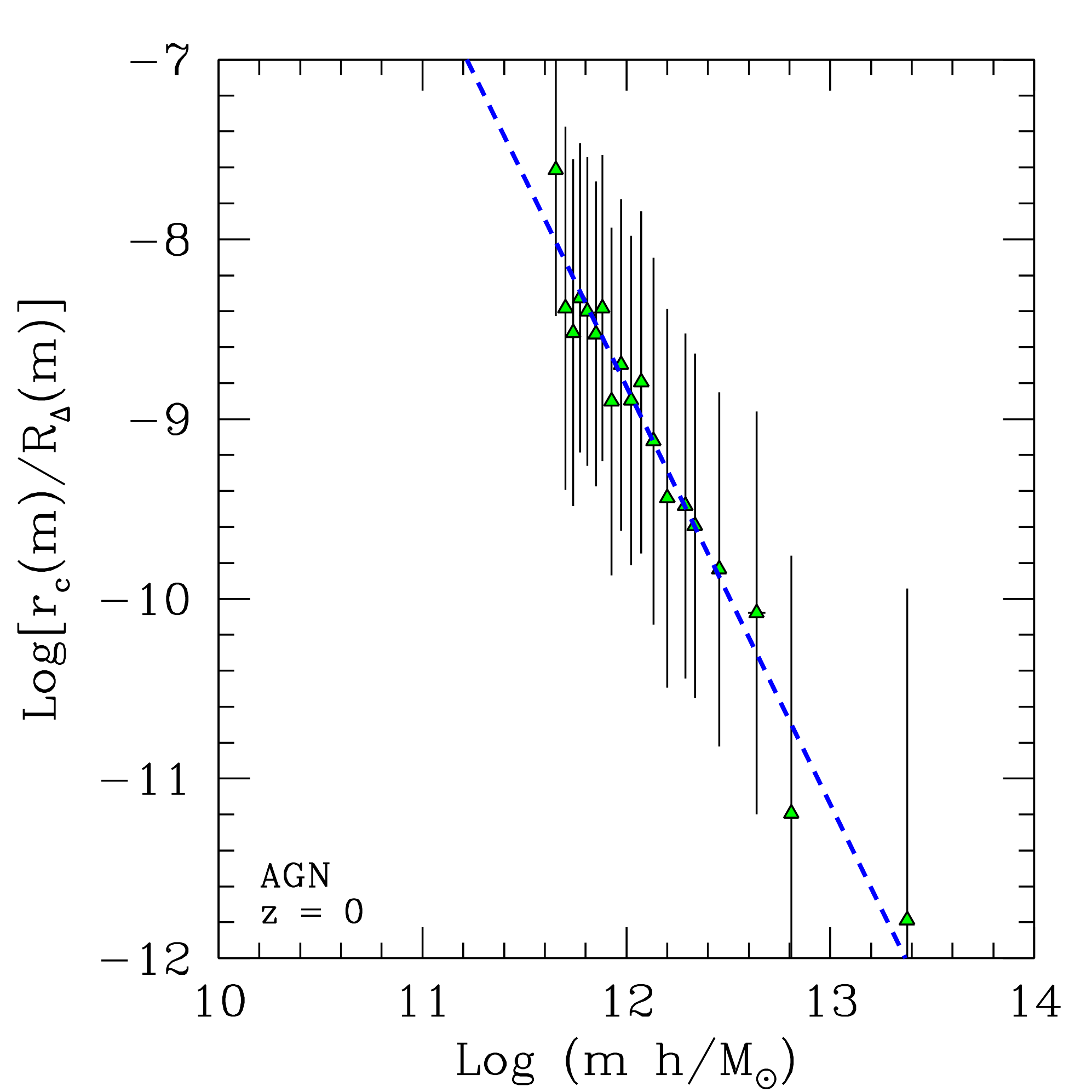}
	\includegraphics[width=0.49\hsize]{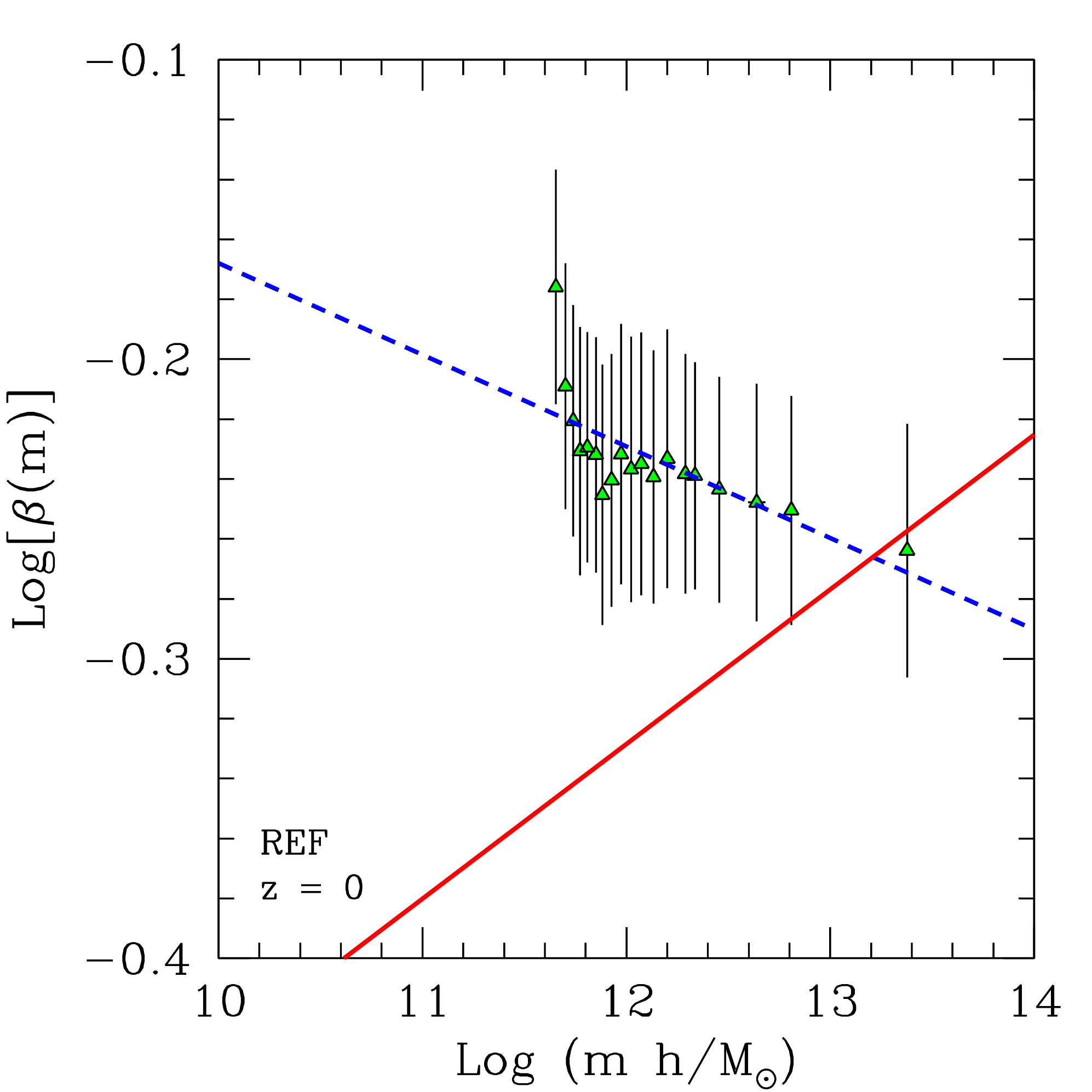}
	\includegraphics[width=0.49\hsize]{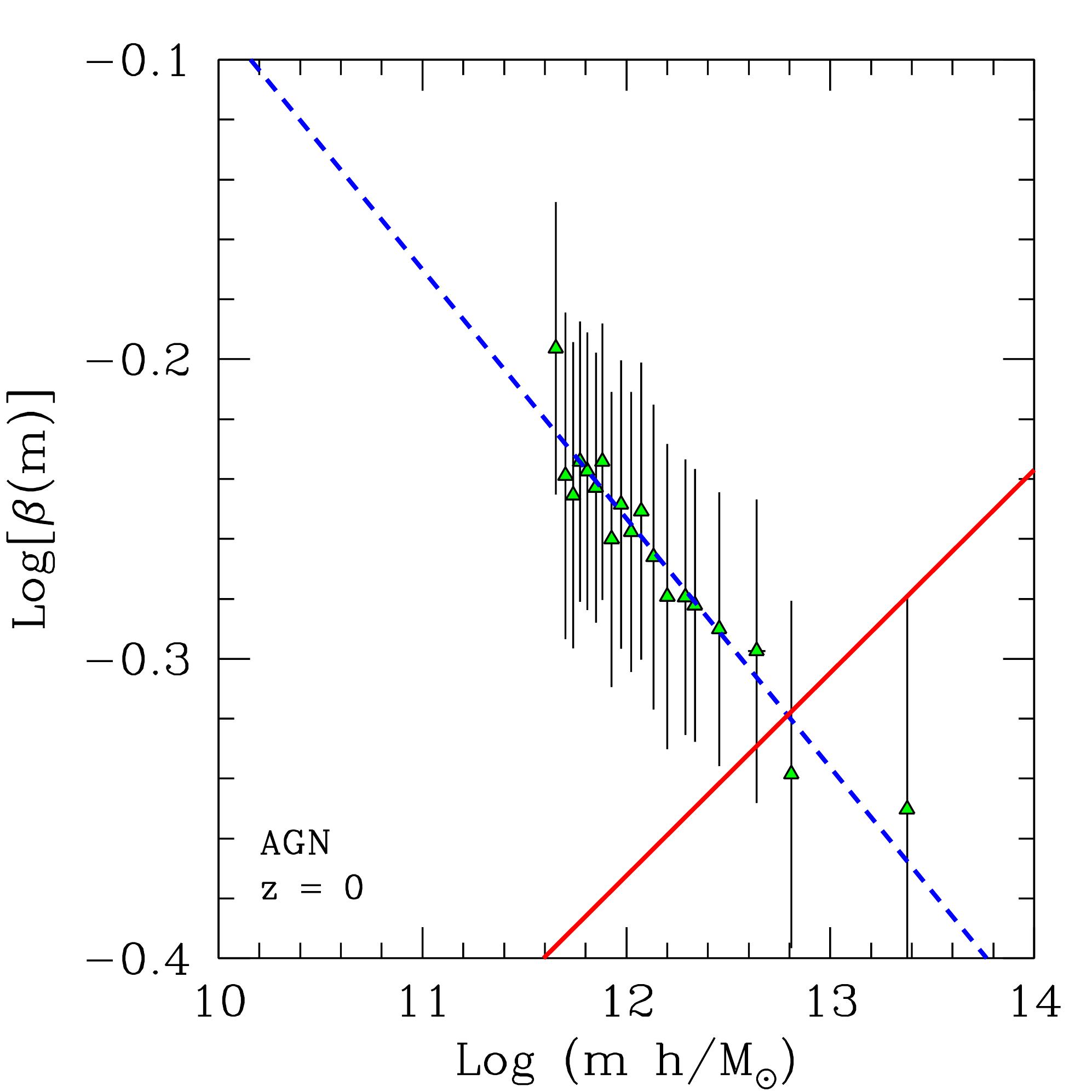}
	\caption{\emph{Upper row}. The ratio of the gas core radius to the equivalent virial radius as a function of the equivalent mass. Triangles represent results for the REF run (left panel) and AGN run (right panel). Halos have been stacked within mass bins so that about the same number is present within each bin. Vertical bars show the scatter around the mean stacked profiles. The blue dashed lines show the best fit power-laws. \emph{Lower row}. As the upper row, but showing the outer slope $\beta$ of the gas density profile as a function of equivalent mass. The solid red lines show the best fit to the gaseous power spectrum measured in the OWLS simulations.}
	\label{fig:gas}
\end{figure}

We proceed in a similar way for the clustering of hot gas. This baryonic component has been modeled via a $\beta-$profile:

\begin{equation}
\rho_\mathrm{g}(r|m) = \frac{\rho_\mathrm{c}}{\left[1+(r/r_\mathrm{c})^2\right]^{3\beta/2}}\;,
\end{equation}
which has two free parameters: the core radius $r_\mathrm{c}$ and the outer slope $\beta$ (the core density $\rho_\mathrm{c}$ is constrained by the overall gas mass fraction). In Figure \ref{fig:gas} we show the mass dependence of these two parameters for the REF (left panels) and AGN (right panels) simulations. It turns out that the core radius is always a very small fraction of the equivalent virial radius in the mass range that is probed by the OWLS simulations considered here. This holds for both the REF and AGN runs. At most, $r_\mathrm{c}$ reaches $\sim 1\%$ of $R_\Delta$ for $m\sim 10^{9}h^{-1}M_\odot$, where the gas fraction is already largely negligible (see Figure \ref{fig:fractions}). This is probably due to a cold gas component which becomes relevant on small scales and is not captured by a simple $\beta$ model, see the discussion further below. As a consequence, for all practical purposes the gas density profile can be considered to be a power-law, and the details of how the core radius depend on mass are effectively unimportant. This also means that the numbers displayed in the upper row of Figure \ref{fig:gas} have no direct physical meaning. Nevertheless, we still fitted their mass dependences and fed them to the SAM, in order to maintain self-consistency with the rest of the work. The only relevant parameter thus remains the outer slope of the profile, which is interestingly close to its fiducial value of $\beta = 2/3$ at the low-mass end of the range probed by simulations. By increasing the structure's mass the value of $\beta$ tends to decrease, implying that the hot gas density profiles become flatter as more massive structures are considered. Also, gas distributions within the structures of the AGN run are on average flatter than in the REF run. This can be expected, because the strong energy feedback present in the latter simulation causes large amounts of gas to be redistributed away from the centers of structures.

\begin{figure}
	\centering
	\includegraphics[width=0.49\hsize]{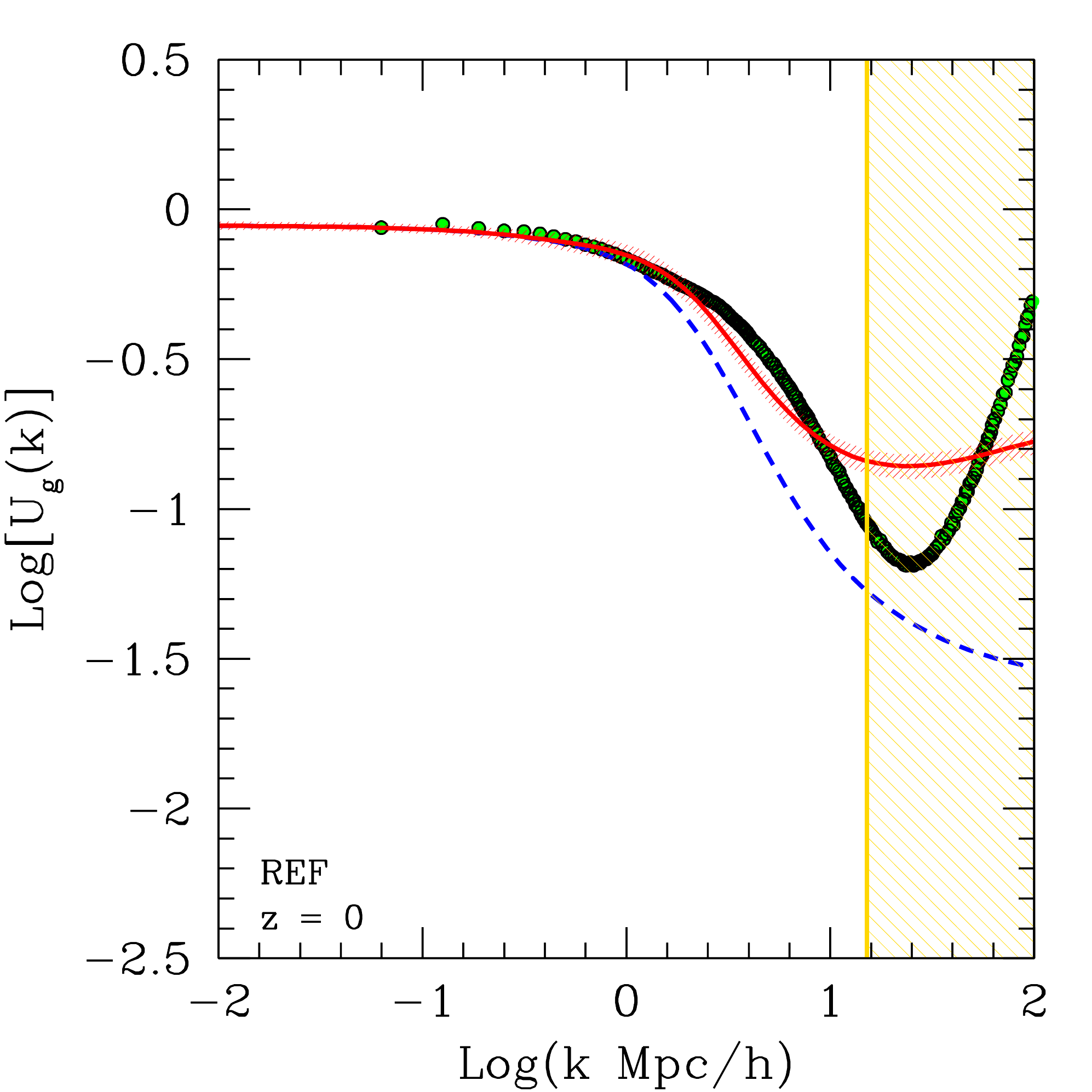}
	\includegraphics[width=0.49\hsize]{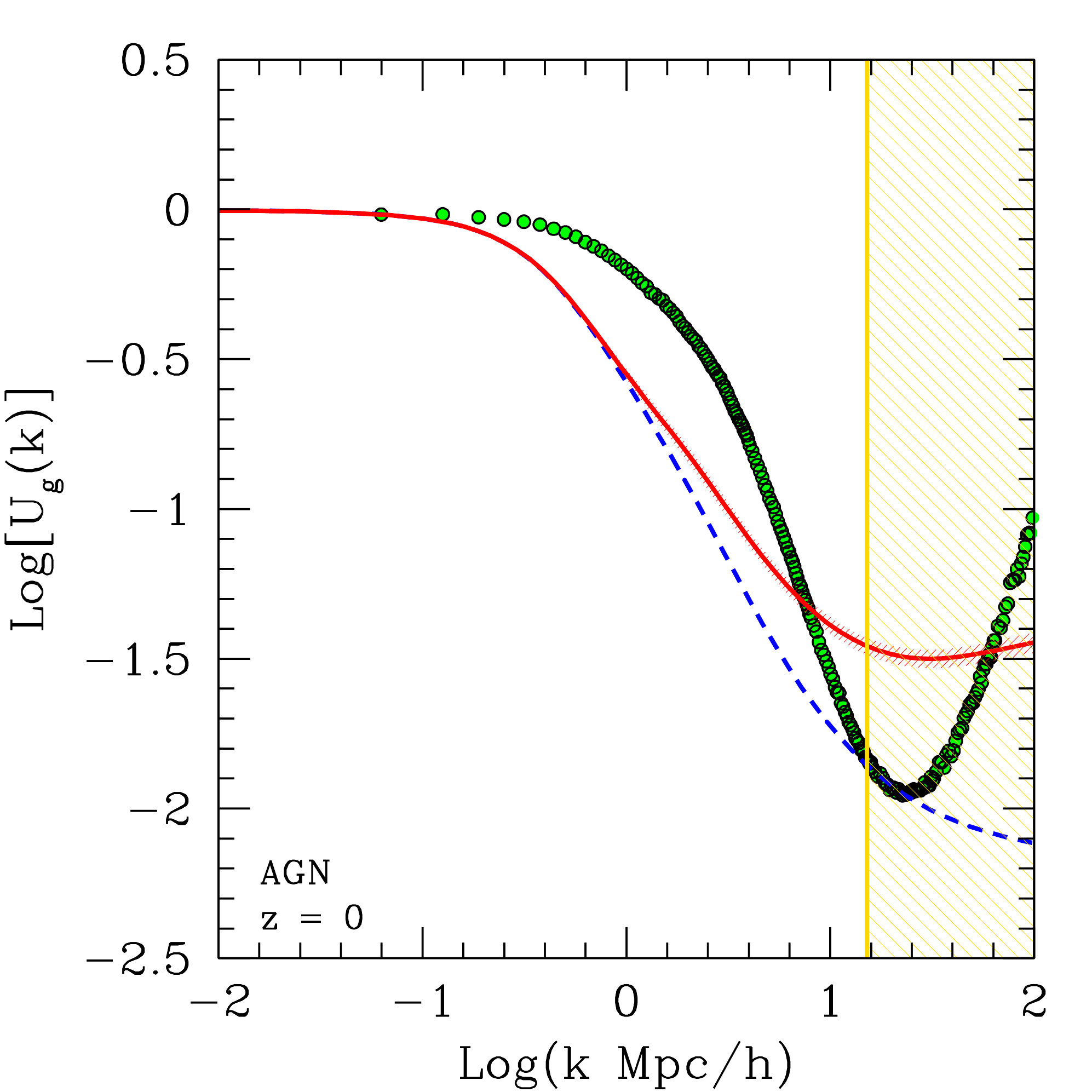}
	\caption{The ratio of the gas power spectra in the REF (left panel) and AGN (right panel) runs to the DM power spectrum in the DMONLY simulation (green circles). The blue dashed lines show the gas power spectra computed with the SAM adopting the best fit parameters for the gas profiles shown in Figure \ref{fig:gas}. The red solid lines show the best fit power spectra obtained by tuning the value of $\beta$ as a function of mass, while the red shaded regions display the effect of a $\pm 10\%$ uncertainty on the theoretical halo mass function. The yellow regions represent the range of scales that has not been included in the fits.}
	\label{fig:gasPower_U_COMPARISON}
\end{figure}

In Figure \ref{fig:gasPower_U_COMPARISON} we show the gas power spectra measured in the REF and AGN runs, compared to the results of the SAM. As mentioned in Paper I, the bias of the diffuse gas component, which dominates the clustering on large scales, is a free parameter that needs to be calibrated against the simulations. We performed this calibration for both the REF and AGN runs, resulting in values of $b_\mathrm{d} = 0.85$ and $b_\mathrm{d} = 0.99$, respectively. According to the recent observational results by \cite{VA13.3} (see also \cite{MA14.1}) these bias values correspond to temperatures for the diffuse gas component of $\sim 2-2.4$ keV for an electron number density of $\sim 1$ m$^{-3}$. Interestingly, the clustering of the diffuse gas component is $\sim 35\%$ lower in the REF simulation than in the AGN one, despite the fact that in the latter the higher energy injection should tend to make the gas distribution more uniform. This is probably due to the fact that the AGN feedback increases the gas fraction in massive objects (where less gas gets converted into stars with respect to the case in which AGN are ignored), and this gives a higher weight to the large-scale clustering contribution of individual halos. On the other hand, on small scales the clustering of gas (which is dominated by the mutual correlation between particles belonging to the same structure) gets much more substantially suppressed in the AGN run than in the REF run. At $k\sim 15h$ Mpc$^{-1}$ the reduction factor of the gaseous power spectrum with respect to the DM spectrum in the baryon-free case is $\sim 15$ in the REF case, while it reaches $\sim 100$ in the AGN case. This is clearly a consequence of the stronger feedback in the second simulation with respect to the first one.

The gaseous power spectrum returned by the SAM with the best fit values of $\beta$ depicted by the blue dashed lines in Figure \ref{fig:gas} tracks this qualitative behavior for $k\lesssim 15h$ Mpc$^{-1}$, implying that the basic physical ingredients are encapsulated by the halo model. On smaller scales the simulated power spectra of gas tend to rise back toward the DMONLY power spectrum, a trend that interestingly is not captured by the SAM. We interpret this feature as due to the presence of cold gas in the central regions of massive galaxies. When measuring the gas power spectrum in the simulations, no distinction is made based on the temperature, implying that the simulated spectrum includes both cold and hot gas. On the other hand, the $\beta-$profile that we used to model the gas distribution within structures is appropriate only for the hot component. Such a profile cannot account for a cold gas component that would be significantly more clustered, likely similar to the stellar distribution. Recently Cen \cite{CE13.1} used hydrodynamic simulations to show that the gas budget in galaxy-sized structures is dominated by a cold component ($T < 10^5$ K) in the inner $\sim 50-100 h^{-1}$ kpc, which is exactly the range in scales that could explain the upturn. The abundant presence of cold gas that has not been turned into stars even in massive elliptical galaxies has also been confirmed observationally in recent studies \cite{TH12.1}.

In order to take into account the presence of this cold gas component, the gas density profile should be modified to allow for an extra compact term. Alternatively, one could simply add the cold gas to the stellar distribution, taking care of modifying the stellar mass fraction accordingly. For the time being, given that scales $\gtrsim 15h$ Mpc$^{-1}$ are of little importance for cosmic shear, we decided to leave the gas distribution unchanged and to postpone the resolution of this point to one of our follow-up works. Going back to the regime $k \lesssim 15h$ Mpc$^{-1}$, despite the qualitative agreement, there are substantial quantitative differences between numerical simulations and the results of the halo model. Typically, the SAM underestimates the clustering of gas by factors up to $\sim 3-4$, with this discrepancy extending to larger scales in the AGN run than in the REF run. In order to find out the origin of this discrepancy, we fit the simulated power spectra of gas with the SAM spectra by letting the value of $\beta$ vary freely. In order to avoid the model trying to capture the small-scale upturn, we limited the fitting at $k \le 15h$ Mpc$^{-1}$. The results of this fitting are shown as solid red lines in Figure \ref{fig:gasPower_U_COMPARISON}, with shaded regions displaying the effect of a $\pm 10\%$ uncertainty on the DM halo mass function. The corresponding slope-mass relations are also shown with solid red lines in Figure \ref{fig:gas}.

As can be seen, the agreement with the REF numerical simulation improves substantially in the range of scales that we considered, with the maximum discrepancy now being reduced to $\sim 25\%$. The same is not true for the AGN run: while the gaseous power spectrum produced by the SAM can be increased, this is not nearly enough to mend the discrepancy with simulated results. Another issue that is worth mentioning, is that in order to find a better fit to numerically simulated spectra, the SAM requires a value of $\beta$ that is monotonically increasing with mass, which is a trend opposite to the one observed for individual structures. The fact that in the AGN case even changing drastically the gas density profile does not improve significantly the agreement of the SAM with the numerical result implies a very important result: the bias of the diffuse gas component with respect to the linear DM clustering must have a scale dependence. Specifically, it must increase with increasing spatial frequency. This is the only way in which the discrepancy at $k \lesssim 1 h$ Mpc$^{-1}$ can be resolved, because the diffuse gas component is the next-to-leading contribution to the gas clustering at those scales (see Paper I). Note that extending the halo gas distribution out to several times the equivalent virial radius would effectively move gas from the diffuse component into the halo component, and thus will have an effect similar to a scale dependent diffuse bias. See the recent work in ref. \cite{MA14.1} for such an approach to a related study.

This result is important because it could not be derived by looking at numerical simulations alone (which do not distinguish different gas components) and because it has significant implications for future observations of the diffuse component (which is nowadays only tentatively detected, see e.g., \cite{VA13.3}). At the same time, while the scale dependence of the diffuse component affects also the gas power spectrum in the REF case, it is not as significant there, likely because the lower level of energy feedback allows even low-mass DM halos to retain more significant amounts of gas, and thus reduces the relative contribution of the diffuse component itself at $k \gtrsim 1h$ Mpc$^{-1}$. In fact, we are able to recover the simulated gaseous spectrum by acting only on the individual average density profiles.

The fact that the best fit $\beta$ values as a function of mass are in contradiction with the trends measured in the REF simulation has been traced back to the gas mass fraction. By increasing this fraction for masses larger than those probed by the simulations in Figure \ref{fig:fractions} one raises the contribution of the $1-$halo term to the gas power spectrum, and thus reduces the need for a steep gas profile (large $\beta$ values) at high-masses. Note that this does not help as much in the AGN case, as the discrepancy extends to much larger scales there. Assuming that one has access to more complete information about the baryon fraction, one should in principle also take into account the scale dependence of the diffuse gas bias. This is a straightforward modification of the SAM implementation, however it would mean the addition of $2-3$ new parameters to the model, thus complicating it considerably. We therefore suggest this modification to be implemented only if the current implementation is unable to reproduce some feature of the observed total matter power spectrum or the total matter power spectrum that one aims at reproducing (see Section \ref{sct:total} below), or if in the future some more accurate observation of the diffuse gas component is performed.

\begin{figure}
	\centering
	\includegraphics[width=0.49\hsize]{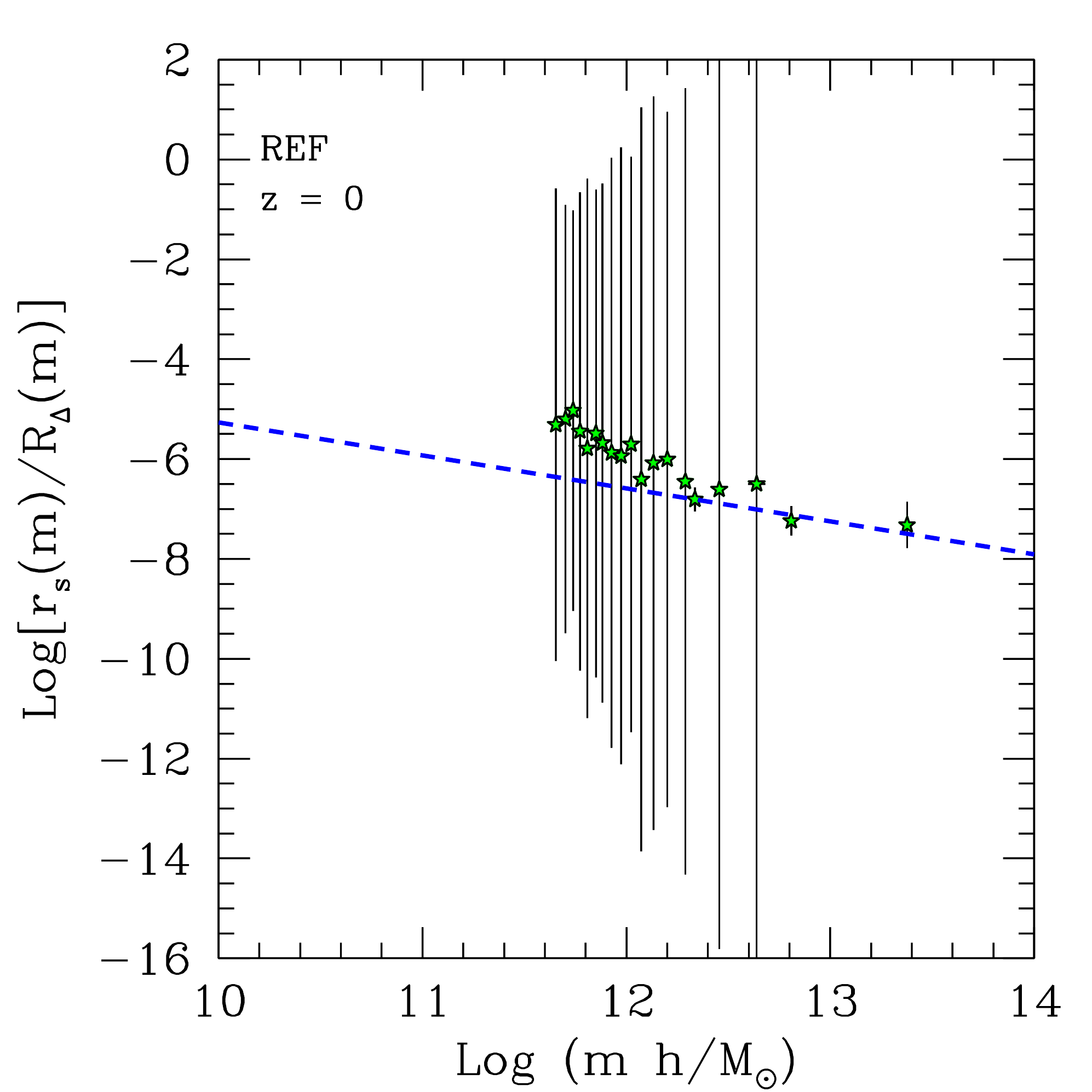}
	\includegraphics[width=0.49\hsize]{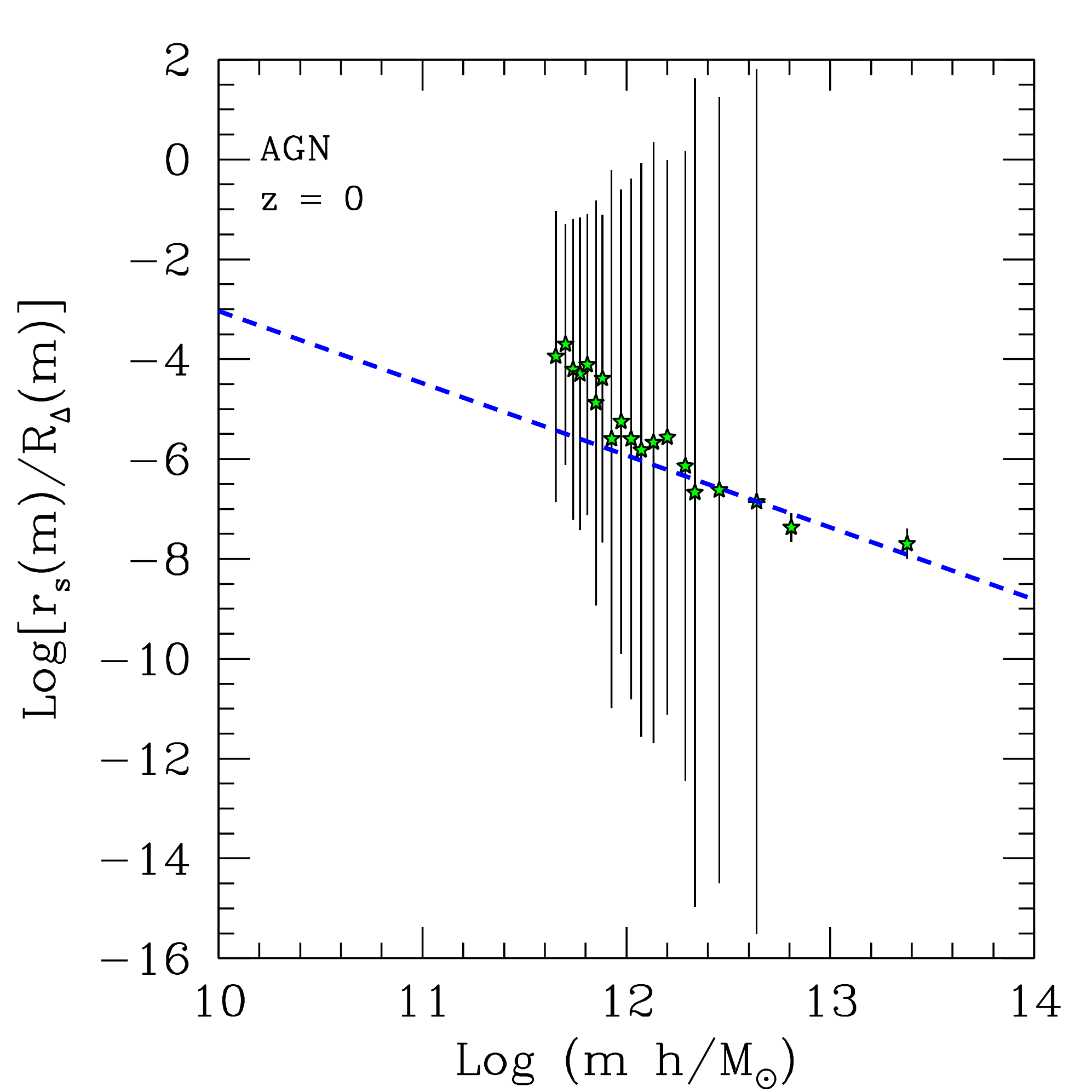}
	\includegraphics[width=0.49\hsize]{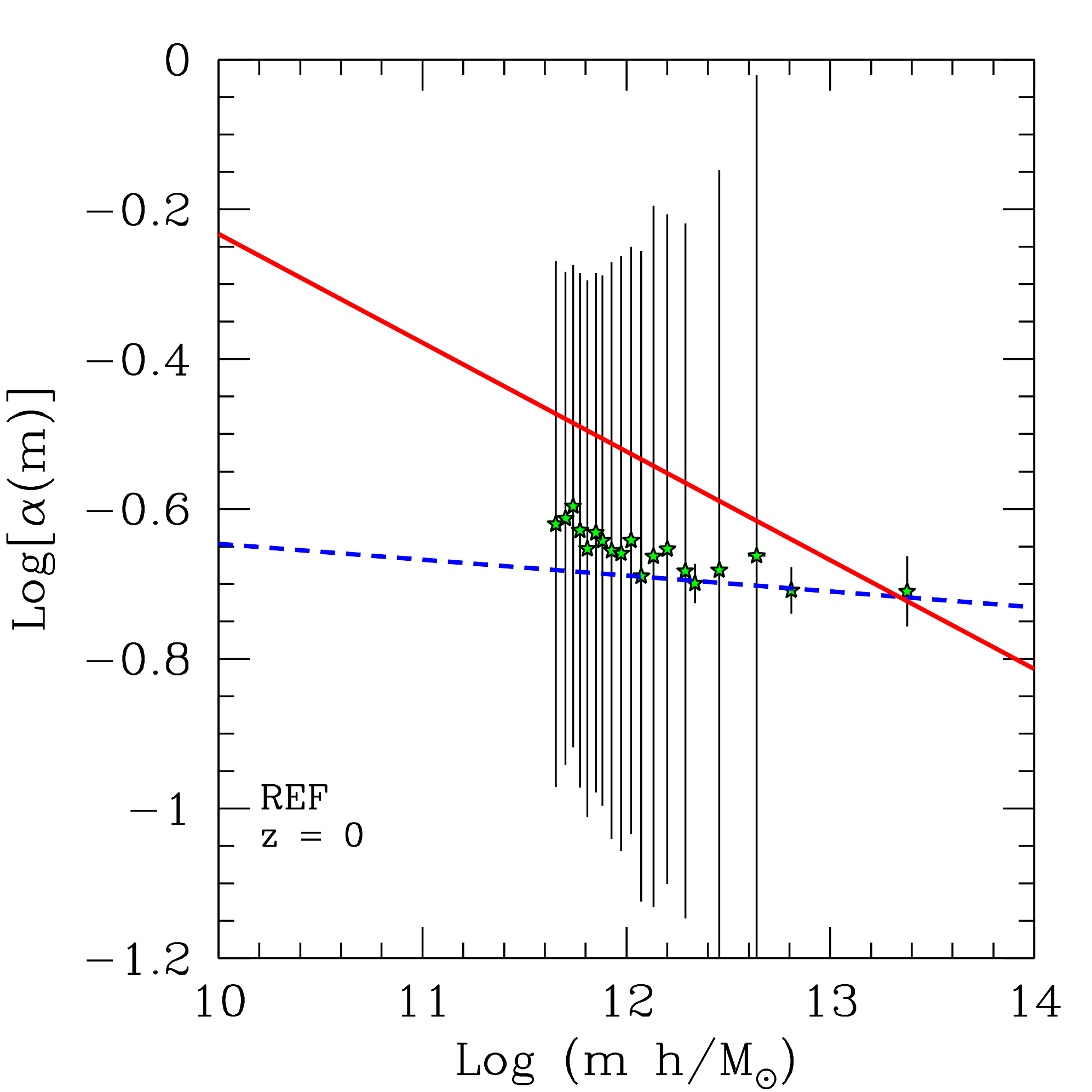}
	\includegraphics[width=0.49\hsize]{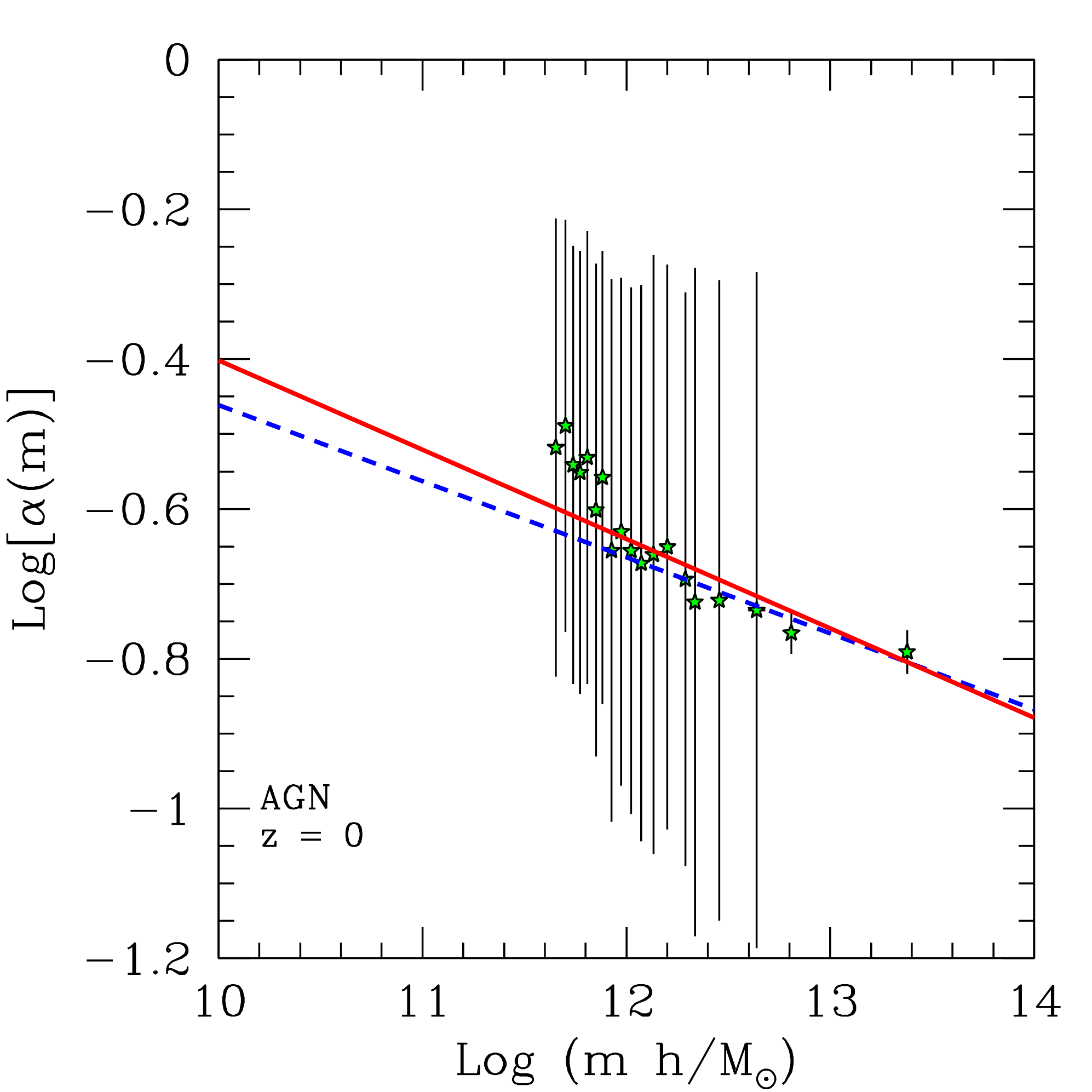}
	\caption{\emph{Top panels}. The ratio of the transition radius for the stellar density profile to the equivalent virial radius as a function of structure equivalent mass. \emph{Bottom panels}. The outer slope of the stellar density profiles as a function of the equivalent mass. The left panels refer to the REF run, while the right panels refer to the AGN run. The dashed blue lines show the best fit power-laws, while the solid red lines correspond to the stellar profiles that better capture the stellar power spectra measured in the simulations.}
	\label{fig:stellar_rs}
\end{figure}

\subsection{Stars}\label{sct:stars}

Finally, we assess the distribution of stars. The stellar distribution within DM halos has been modeled via a modified exponential profile:

\begin{equation}
\rho_\star(r|m) = \frac{\rho_\mathrm{t}}{r/r_\mathrm{t}}\exp\left[ -\left(\frac{r}{r_\mathrm{t}}\right)^\alpha \right]\;,
\end{equation}
which has two independent parameters: the transition radius $r_\mathrm{t}$ and the outer slope $\alpha$ (the transition density $\rho_\mathrm{t}$ is constrained by the overall stellar mass fraction). In the top two panels of Figure \ref{fig:stellar_rs} we show how the transition radius depends on the structure equivalent mass for the REF (left panel) and AGN (right panel) simulations. Much like the core radius for the stacked gas profile, in all cases the transition radius is a very small fraction of the equivalent virial radius, reaching a value of $\sim 0.01\;R_\Delta$ only for $m\sim 5\times 10^9 h^{-1}M_\odot$ in the AGN run. As such, the details of how $r_\mathrm{t}$ depend on mass are physically unimportant. We also note that the scatter within each individual mass bin is very large on galaxy scales, covering many orders of magnitude, while it gets sharply reduced on group scales. This might indicate that the profile we chose does not very accurately describe galactic stellar distributions, while it becomes more satisfactory when satellites and intra-halo stars become relevant.

A similar trend is detected also in the slope $\alpha$ of the profiles, shown in the bottom two panels of Figure \ref{fig:stellar_rs}, although in this case the scatter is only of a factor $\sim 5-6$. We also observe that in both the REF and the AGN simulation the stellar density profiles tend to become shallower as mass increases, which is something one might expect since the distribution of satellite galaxies and of intra-halo stars is much less compact than the stellar distribution within individual galaxies. Despite all of this, we show below that the stellar power spectrum of numerical simulations is quite well reproduced, and thus we find it unnecessary to add further complications to the distribution of stars within individual structures.

\begin{figure}
	\centering
	\includegraphics[width=0.49\hsize]{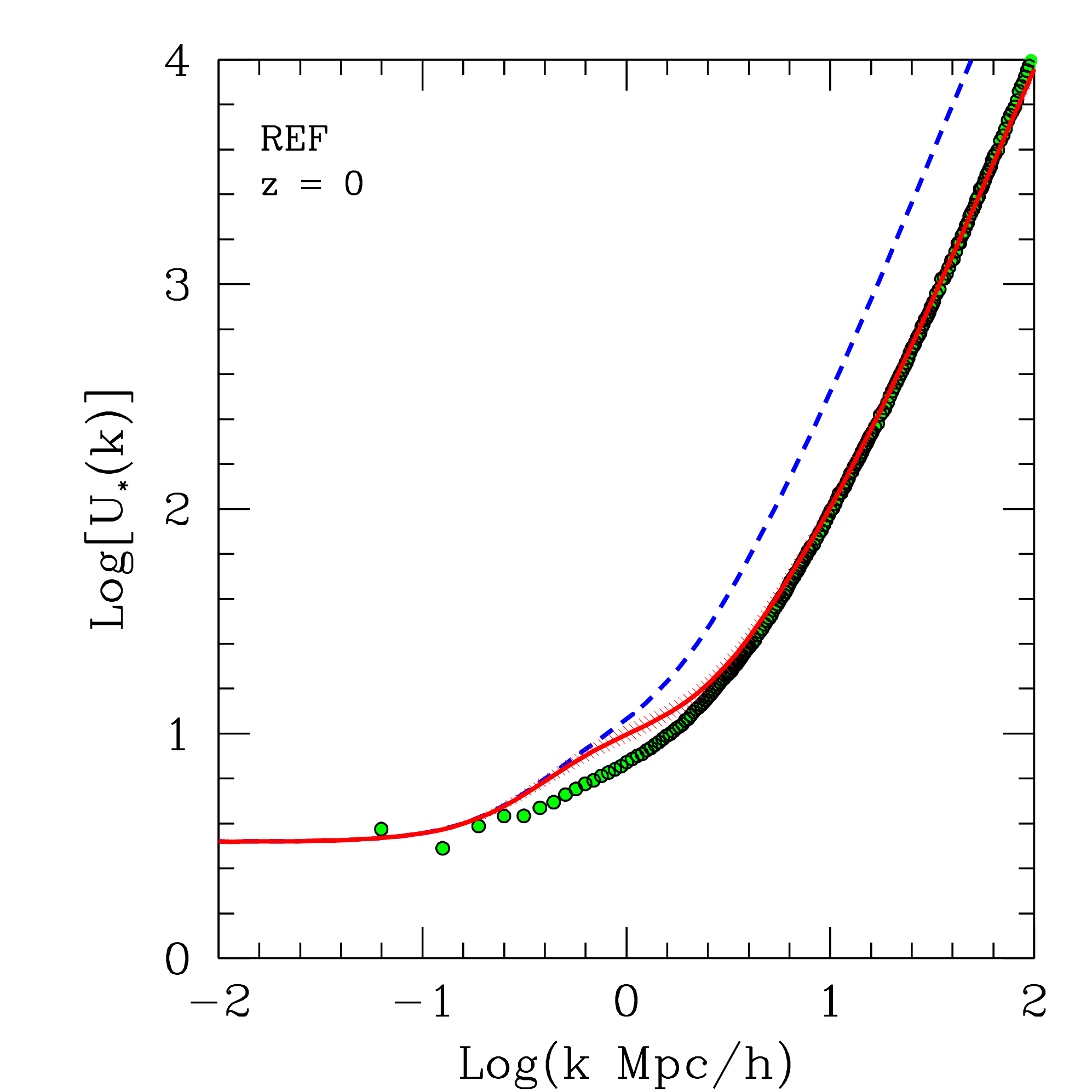}
	\includegraphics[width=0.49\hsize]{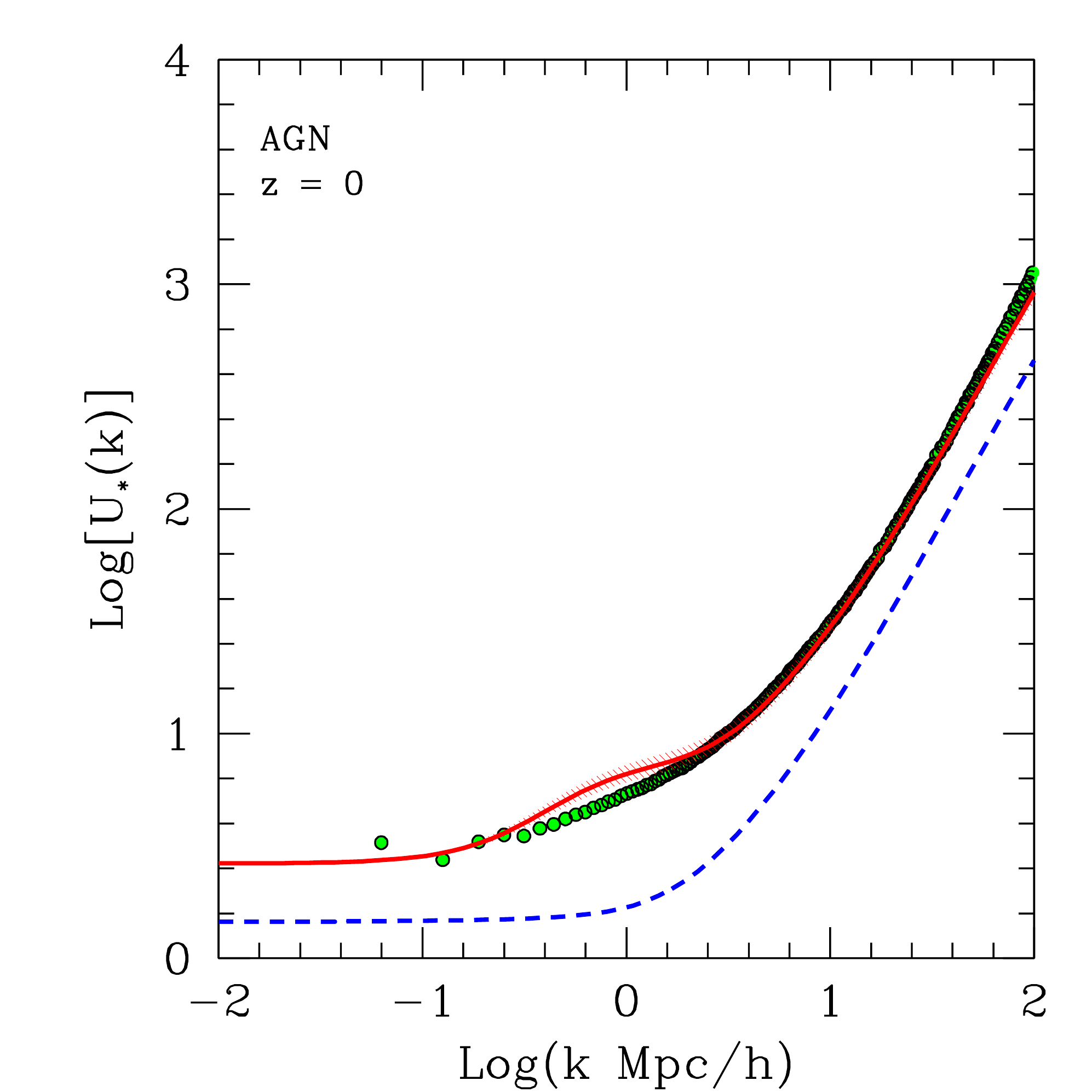}
	\caption{The ratio of the stellar power spectra in the REF (left panel) and AGN (right panel) runs to the DM power spectrum in the DMONLY simulation (green circels). The blue dashed lines show the stellar power spectra computed with the SAM adopting the best fit parameters for the stellar profiles shown in Figure \ref{fig:gas}. The red solid lines show the best fitting spectra obtained by varying the outer slope $\alpha$ of the stellar profiles, while the red shaded regions display the effect of a $\pm 10\%$ uncertainty on the DM halo mass function.}
	\label{fig:stellarPower_U_COMPARISON}
\end{figure}

In Figure \ref{fig:stellarPower_U_COMPARISON} we used green circles to show the stellar power spectra for the REF and AGN runs. It is a well known fact that stars are much more clustered than DM: at small scales this happens because the stellar distribution is substantially more compact than the DM one; at large scales it happens because stars can form only inside sufficiently massive DM halos, which are significantly more clustered than DM itself. While in this latter regime there is no relevant difference between the two hydro simulations considered here, the stellar power spectrum in the AGN run is $\sim 1$ order of magnitude lower than in the REF run at small scales. As mentioned above, this is in line with the stronger energy feedback provided by AGN. Qualitatively this trend is captured by the SAM when using the best fit stellar distribution parameters shown in Figure \ref{fig:stellar_rs}, but there are a number of quantitative differences. Specifically, the stellar power spectrum computed by the SAM overestimates the REF simulation result, more so at small scales. The discrepancy reaches a factor of $\sim 5$ for $k\gtrsim 40h$ Mpc$^{-1}$. On the other hand, the stellar power spectrum measured in the AGN simulation is consistently underestimated by the halo model, also (and more disturbingly) on large scales. This latter fact in particular deserves some contemplation. The large scale discrepancy observed in the AGN panel of Figure \ref{fig:stellarPower_U_COMPARISON} cannot be due to uncertainties in the theoretical DM halo mass function: if it was, then one should also observe a discrepancy in the REF case; plus, the investigation reported in Paper I shows that it is very difficult for this theoretical uncertainty to cover a factor of $\sim 2$.

Rather, this discrepancy is a probably related to the stellar mass fraction. Indeed, the stellar mass fraction is very flat in the mass range probed by the AGN simulation (see Figure \ref{fig:fractions}), so that it is difficult to identify a trend that can help us extrapolate its behavior outside this range. We verified that by reducing the stellar fraction at masses lower than the resolution limit of the simulations (thus increasing the stellar effective bias), brings the large-scale stellar power spectrum back in agreement with the AGN simulation. This procedure agrees with the fact that only few real halos with $\mathrm{Log}(m\;h/M_\odot)\lesssim 9$ are expected to host galaxies \cite{SA14.1}. Thus, when fitting the stellar power spectrum of the AGN simulation we first performed this correction to the stellar fraction. The solid red lines show the best fit SAM results by changing the outer slopes of the stellar profiles $\alpha$. The corresponding power-laws are also displayed as solid red lines in the bottom panels of Figure \ref{fig:stellar_rs}, and are seen to be steeper than the original best fits, while still being consistent within the scatter. The results are in excellent agreement with the simulated stellar power spectra at all scales, with the only exception of a $40\%$ overestimation for $k\sim 0.2h$ Mpc$^{-1}$ for both simulations. This discrepancy should be related to a shortcoming of the model at the transition between the $1-$halo and the $2-$halo contributions, possibly still due to the uncertainties in the stellar mass fractions at high masses. However, we do not investigate it further here because the contribution of the stellar clustering to the total matter power spectrum is negligible at those scales (see Paper I).

\subsection{Total matter power spectrum}\label{sct:total}

\begin{figure}
	\centering
	\includegraphics[width=0.49\hsize]{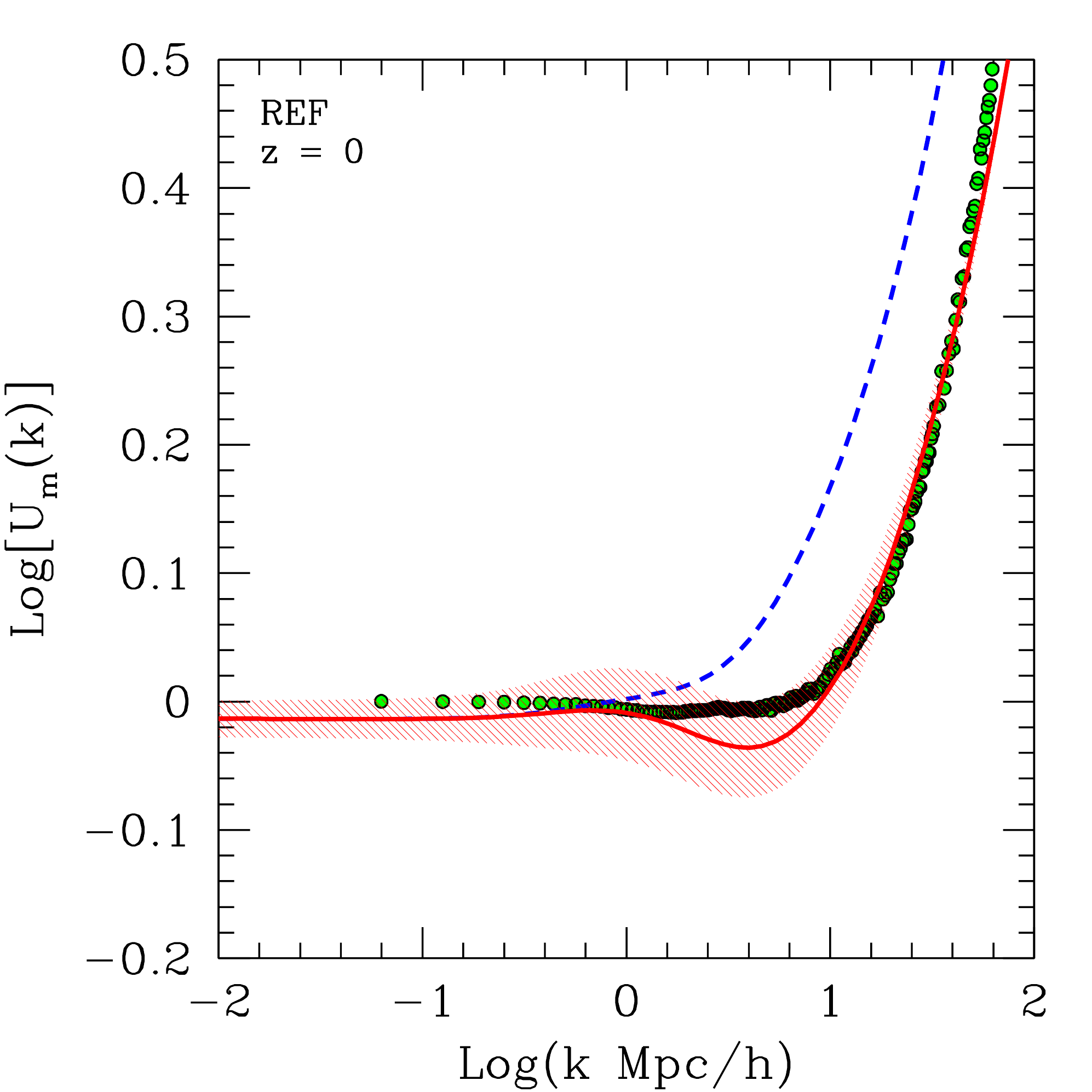}
	\includegraphics[width=0.49\hsize]{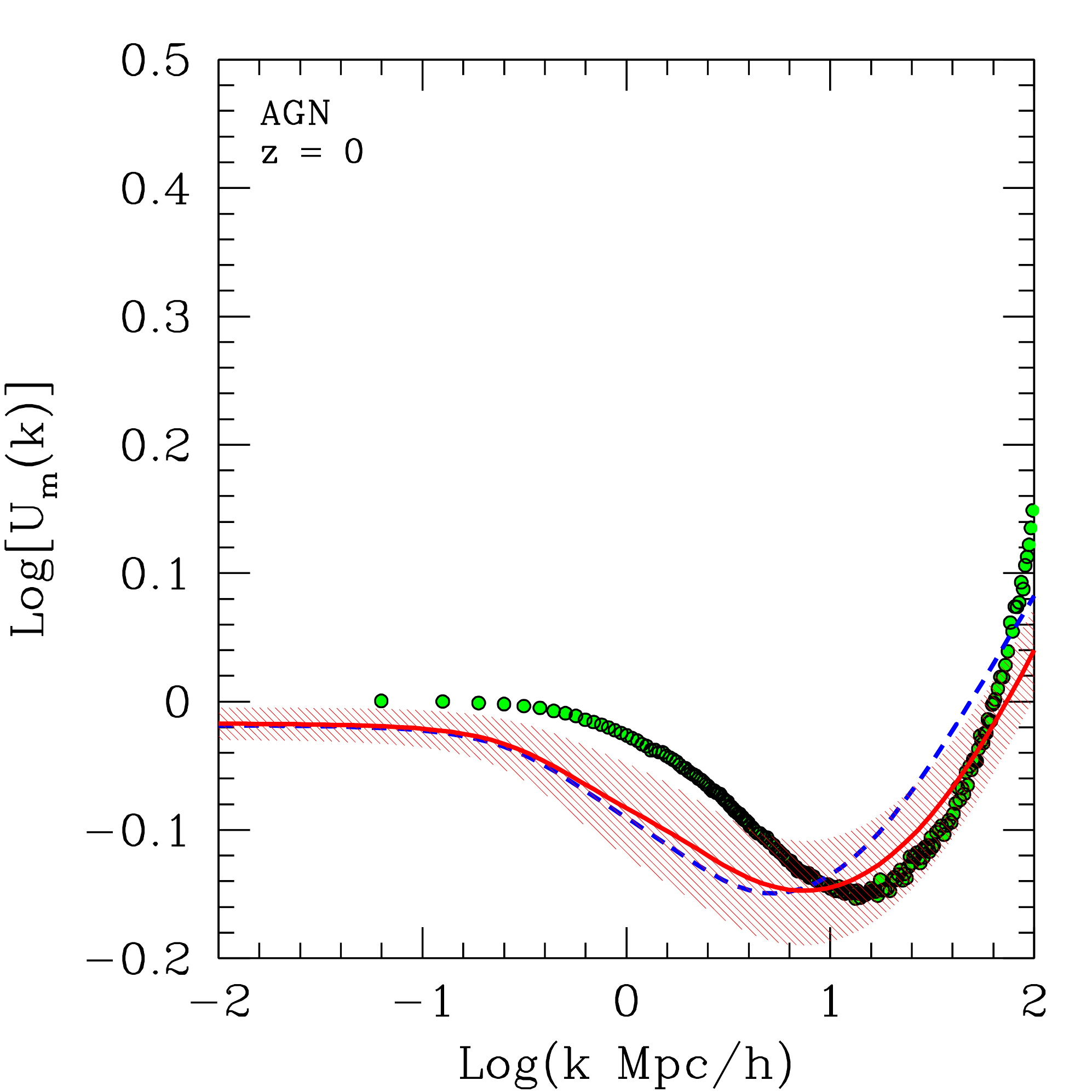}
	\caption{The ratio of the total matter power spectra in the REF (left panel) and AGN (right panel) runs to the DM power spectrum in the DMONLY simulation (green circels). The blue dashed lines show the matter power spectra computed with the SAM adopting the best fit parameters for the stellar, gas, and DM profiles. The red solid one show the result of adopting the best fitting parameters for the clustering of each individual component, and the red dashed region represents the effect of a $\pm 10\%$ uncertainty on the theoretical DM halo mass function.}
	\label{fig:matterPower_U_COMPARISON}
\end{figure}

As a final step, we combined the power spectra of each mass component and their mutual cross spectra to produce the \emph{total} mass power spectrum. We did this by using either the parameters that best fit the matter density profiles within individual halos or the parameters that best fit the clustering of individual matter components. The results are displayed in Figure \ref{fig:matterPower_U_COMPARISON}, where the green circles represent the simulated matter power spectra for both the REF and AGN cosmological runs. As expected, the total matter clustering as a function of scale looks quite similar to the DM clustering, because the DM contribution is always the dominant one. There are, however, significant differences. For instance, the presence of gas tends to decrease the matter power on intermediate scales ($k \sim 1-10h$ Mpc$^{-1}$), especially in the AGN run where gas is particularly modestly clustered. At the same time, the total matter power spectrum is pushed up at small scales compared to the DM, because of the presence of strongly clustered stars.

The blue dashed curves in Figure \ref{fig:matterPower_U_COMPARISON} show the SAM results by assuming the best fitting parameters for individual matter density profiles. As can be seen, the matter power in the REF simulation is always abundantly overestimated for $k \gtrsim 1 h$ Mpc$^{-1}$, by up to $50-100\%$. This is due to the fact that the DM power and the stellar power are also overestimated, for reasons that we discussed in the previous Sections. The fact that gas clustering is instead underestimated by the SAM cannot counteract this. The situation is a bit more complicated for the AGN simulation. There the total matter power is underestimated on intermediate scales and overestimated on small scales. This latter feature is due to the fact that at small scales the DM power is also overestimated, and so is (as in the REF run) the stellar power. The power underestimation at intermediate scales instead stems from the fact that the gas power is largely underestimated by the SAM on those scales (recall that the SAM gas power falls much shorter of the simulated one in the AGN than in the REF simulation, and that this happens also at substantially larger scales, where the gas contribution is more important).

By using the model parameters that best fit the clustering of individual mass components the situation improves visibly in the REF case. As a matter of fact, the total matter power spectrum is accurately captured within the theoretical uncertainties due to a $\pm 10\%$ shift of the DM halo mass function. There is the tendency for the SAM power spectrum to underestimate (by $\sim 5\%$) the simulated one on scales $1h$ Mpc$^{-1} \lesssim k \lesssim 10 h$ Mpc$^{-1}$, which is due to the fact that the gas power spectrum is not accurately captured by the model on those scales. However, this happens to be well within the uncertainties of the model. The improvement is much less evident in the AGN case. Specifically, while for $k\gtrsim 5h$ Mpc$^{-1}$ the halo model works rather well, on smaller spatial frequencies the underestimation of power that we note above persists. This happens because the gas power spectrum measured in the AGN simulation cannot be reproduced even by changing the gas density profiles at will. Thus, this is a situation where the scale dependence of the diffuse component bias would become important also for the total matter clustering. This very important conclusion can be rephrased as follows: if the real Universe looks like the AGN simulation, by using high precision cosmic shear observations in conjunction with the SAM developed in Paper I we can gather information about the clustering of the diffuse gas component.

In conclusion, we note an additional feature in Figure \ref{fig:matterPower_U_COMPARISON}, namely that the large-scale ($\mathrm{Log}(k\;\mathrm{Mpc}\;h^{-1}) < -0.5$) matter power measured in the AGN simulation is not accurately reproduced by the SAM, not even assuming a $\pm 10\%$ theoretical uncertainty on the DM halo mass function. This cannot be due to the clustering of any individual mass component, as we have shown in the previous Sections that their behavior at small spatial frequencies is well captured. Nor can it be due to a misestimation of the abundance of some component: the stellar mass fraction is normalized so as to reproduce the average stellar density (measured in the simulations); the gas density is such that its addition to the stellar density returns the total baryon density, which is an input of the simulations; and the DM density is also an input of the simulations. The only explanation is thus a higher uncertainty in the DM halo mass function and/or the uncertainty in the DM halo linear bias. Indeed, despite being subdominant, the latter can have an effect of a few percent on the total matter clustering at large scales. This would also explain why we do not see such a discrepancy in the clustering of individual components. As discussed in Paper I, the DM and stellar clustering are normalized in such a way that a constant shift in the halo mass function produces no effect, while the effect on the gas clustering is effectively absorbed in the bias of the diffuse component.

\section{Discussion and conclusions}\label{sct:conclusions}

The aim of this paper was to apply the SAM developed in Paper I to a set of numerical hydro simulations, in order to understand the accuracy of the model approximations and to illustrate the complementarity of the model with respect to cosmological runs. The main idea was to display how the SAM can be used to gain new insight into the physics of structure formation, and how it can be improved in the future for this purpose. For the purpose of comparison, we considered two simulations belonging to the OWLS project: the REF run, which includes radiative gas cooling, star formation, supernova feedback and metal enrichment; and the AGN run, which includes all the physics of the REF run with the addition of AGN-powered winds. The main messages that can be taken from this work are summarized as follows.

\begin{itemize}
\item The power spectra of DM and stars measured in the simulations can be reproduced with high accuracy by the SAM with simple mass fraction and density profile parametrizations.
\item In the case of DM, the concentration-mass relation that best reproduces the DM power spectrum measured in the REF simulation is markedly different from the same relation measured directly from individual DM halos, although the two are still compatible given the large scatter in the latter.
\item The stellar mass fraction of structures with $m \lesssim 4\times 10^{11}h^{-1} M_\odot$ that is needed to reproduce the simulated clustering of stars measured at large scales in the AGN run is also somewhat different from the extrapolation of the same fraction from direct measurements within virialized structures. Indeed, the measured stellar mass fraction turns out to be rather flat within the small mass range directly accessible to simulations, thus making the extrapolation process uncertain.
\item The clustering of gas is more difficult to interpret. Specifically, gas has a cold and highly clustered component that is not taken into account in the model and that causes the gas power spectrum to rise significantly at small scales, $k \gtrsim 20h$ Mpc$^{-1}$. This component has no substantial effect on the total mass power spectrum, especially at the scales relevant for cosmic shear. However it is a point for possible future improvement.
\item Apart from this, the gas power spectrum is also difficult to reconcile with its simulated counterpart. This is a combination of the limits on the measured gas mass fractions and, more importantly, the scale dependence of the bias of the diffuse component (the Warm-Hot Intergalactic Medium).
\end{itemize}

The scale dependence mentioned in the last point has to produce an increase in the clustering power of the diffuse gas component as a function of spatial frequency for $k \gtrsim 0.1h$ Mpc$^{-1}$, and is particularly evident for the AGN simulation. This is a very important result for future detection prospects of the diffuse gas component (see for instance \cite{VA13.3}), and will also be a sensible improvement to be implemented in the future. From the previous points it also emerges clearly that by using the SAM we can gather information about the average growth of the large-scale structure that are not accessible by just looking at individually resolved objects. For instance, the stellar clustering gives information about the shape of the stellar mass fraction that is complementary to that obtained by direct measurements in separate structures.

Our future aim is to consider high-precision mock cosmic shear observations and explore the space of SAM parameters using Monte Carlo Markov Chains (MCMCs) in the context of Bayesian inference (see \cite{LU11.1,LU12.1} for a similar application to the galaxy luminosity function). This will allow one to understand which parameters are most strongly constrained, the degeneracies between each, and also to what extent these are changed by introducing the improvements to the SAM discussed in this paper (e.g., a more complex characterization of the DM concentration-mass relation, or the scale dependence of the diffuse gas component). This will allow us to use this generalized halo model as a powerful tool in order to explore the distribution of mass in the Universe, and also to forecast the impact of baryonic physics on the estimation of cosmological parameters.

\section*{Acknowledgments}

CF was partially supported by the University of Florida through the Theoretical Astrophysics Fellowship, and has received funding from the European Commission Seventh Framework Programme (FP7/2007-2013) under grant agreement n$^\circ$ 267251. We are grateful to all the members of the OWLS collaboration for their contributions to the simulations presented here. The simulations were run on Stella, the LOFAR BlueGene/L system in Groningen, on the Cosmology Machine at the Institute for Computational Cosmology in Durham (which is part of the DiRAC Facility jointly funded by STFC, the Large Facilities Capital Fund of BIS and Durham University) as part of the Virgo Consortium research programme, and on Darwin in Cambridge. This simulation work was sponsored by the National Computing Facilities Foundation (NCF) for the use of supercomputer facilities, with financial support from the Netherlands Organization for Scientific Research (NWO). We also gratefully acknowledge funding from the European Research Council under the European Union's Seventh Framework Programme (FP7/2007-2013)/ERC Grant agreement n$^\circ$ 278594-GasAroundGalaxies.

{\small
\bibliographystyle{JHEP}
\bibliography{./master}
}

\end{document}